\documentclass[submission,Phys]{SciPost}
\pdfoutput=1
\usepackage{amsmath,amssymb,mathtools,xspace}
\usepackage{booktabs,multirow,graphicx,tabularx,slashed}
\usepackage{hyperref}
\usepackage{color,xcolor}
\usepackage[normalem]{ulem}
\usepackage{enumitem}
\usepackage{feynmp}
\usepackage{braket,stackrel}
\usepackage{tikz}
\usepackage{subcaption}
\usetikzlibrary{arrows}
\usetikzlibrary{shapes.geometric, arrows}

\makeatletter
\@ifundefined{pdfoutput}{}{\DeclareGraphicsRule{*}{mps}{*}{}}
\makeatother

\makeatletter
\DeclareRobustCommand*{\bfseries}{%
   \not@math@alphabet\bfseries\mathbf
   \fontseries\bfdefault\selectfont
   \boldmath
}
\makeatother

\setitemize{itemsep=2pt,topsep=2pt,parsep=0pt,partopsep=0pt,leftmargin=*}
\setenumerate{itemsep=0pt,topsep=2pt,parsep=0pt,partopsep=0pt,labelindent=3pt,leftmargin=*}

\definecolor{Gcolor}{HTML}{3b528b}
\definecolor{Dcolor}{HTML}{e41a1c}

\tikzstyle{generator} = [rectangle, rounded corners, minimum width=3cm, minimum height=1cm,text centered, draw=Gcolor]
\tikzstyle{discriminator} = [rectangle, rounded corners, minimum width=3cm, minimum height=1cm,text centered, draw=Dcolor]
\tikzstyle{io} = [circle, trapezium left angle=70, trapezium right angle=110, minimum width=1cm, minimum height=1cm, text centered, draw=black]

\tikzstyle{process} = [rectangle, minimum width=1cm, minimum height=1cm, text centered, draw=black]
\tikzstyle{decision} = [rectangle, minimum width=1cm, minimum height=1cm, text centered, draw=black]

\tikzstyle{arrow} = [thick,->,>=stealth]
\usepackage{xcolor}

\newcommand\one{\leavevmode\hbox{\small1\normalsize\kern-.33em1}}

\newcommand{\mat}{\mathcal{M}}

\newcommand{\qqquad}{\qquad \qquad}

\newcommand{\gev}{\text{GeV}}

\def\slashchar#1{\setbox0=\hbox{$#1$}           
   \dimen0=\wd0                                 
   \setbox1=\hbox{/} \dimen1=\wd1               
   \ifdim\dimen0>\dimen1                        
      \rlap{\hbox to \dimen0{\hfil/\hfil}}      
      #1                                        
   \else                                        
      \rlap{\hbox to \dimen1{\hfil$#1$\hfil}}   
      /                                         
   \fi}

\newcommand{\fastjet}{\textsc{FastJet}\xspace}
\newcommand{\delphes}{\textsc{Delphes}\xspace}
\newcommand{\sherpa}{\textsc{Sherpa}\xspace}

\begin{document}

\begin{center}{\Large \textbf{
Measuring QCD Splittings with Invertible Networks
}}\end{center}

\begin{center}
Sebastian Bieringer\textsuperscript{1},
Anja Butter\textsuperscript{1},
Theo Heimel\textsuperscript{1},
Stefan H\"oche\textsuperscript{2},
Ullrich K\"othe\textsuperscript{3}, 
Tilman Plehn\textsuperscript{1}, and 
Stefan T. Radev\textsuperscript{4} 
\end{center}

\begin{center}
{\bf 1} Institut f\"ur Theoretische Physik, Universit\"at Heidelberg, Germany\\
{\bf 2} Fermi National Accelerator Laboratory, Batavia, IL, USA \\
{\bf 3} Heidelberg Collaboratory for Image Processing, Universit\"at Heidelberg, Germany\\
{\bf 4} Psychologisches Institut, Universit\"at Heidelberg, Germany \\
heimel@thphys.uni-heidelberg.de
\end{center}

\begin{center}
\today
\end{center}


\tikzstyle{int}=[thick,draw, minimum size=2em]

\section*{Abstract}
 {\bf QCD splittings are among the most fundamental theory concepts at
   the LHC. We show how they can be studied systematically with the
   help of invertible neural networks. These networks work with
   sub-jet information to extract fundamental parameters from jet
   samples. Our approach expands the LEP measurements of QCD Casimirs
   to a systematic test of QCD properties based on low-level jet
   observables. Starting with an toy example we study the effect of
   the full shower, hadronization, and detector effects in detail. }

\vspace{10pt}
\noindent\rule{\textwidth}{1pt}
\tableofcontents\thispagestyle{fancy}
\noindent\rule{\textwidth}{1pt}

\newpage
\section{Introduction}
\label{sec:intro}

The upcoming Run~3 and HL-LHC are starting an era of precision physics
at hadron colliders. With this perspective we need to re-think our
strategies for measurements, interpretation frameworks, and
first-principle theory predictions. A big step in the direction of
Machine Learning (ML) based measurements has been made in jet classification based on low-level
detector output. It starts from the observation that subjet taggers
benefit immensely from multivariate
approaches~\cite{Lonnblad:1990bi,Csabai:1990tg}, combined with the
ability of modern convolutional networks to extract subjet
information~\cite{Cogan:2014oua,deOliveira:2015xxd,Baldi:2016fql,Komiske:2016rsd,Kasieczka:2017nvn,Macaluso:2018tck}. Alternatively,
we can feed a network the subjet
4-momenta~\cite{Almeida:2015jua,Butter:2017cot,Pearkes:2017hku,Erdmann:2018shi},
or change the architecture to recurrent networks~\cite{Louppe:2017ipp}
or point clouds~\cite{Komiske:2018cqr,Qu:2019gqs}. Top tagging is an
especially interesting subjet problem, because the tagger output is
theoretically well defined and the training can be done on data
only. This makes top taggers an excellent conceptual testing
ground~\cite{Kasieczka:2019dbj}, including the crucial question how to
control
uncertainties~\cite{Bollweg:2019skg,Kasieczka:2020vlh}. Already in the
context of taggers, the focus on data-driven ML-applications becomes a
major problem for particle physics when it breaks our central promise
of understanding LHC collisions entirely in terms of fundamental
physics. The question then becomes how we can use these new approaches
to improve our understanding of QCD~\cite{Dreyer:2018nbf,Lai:2020byl}.

From a theory perspective, the LHC objects described by the simplest
fundamental laws, at least to leading order, are parton
showers~\cite{Marchesini:1987cf}. In perturbative QCD their entire
behavior can be described by the quark-gluon interaction and the
triple gluon interaction. Their leading kinematic behavior can be
understood as logarithmically enhanced collinear and soft splittings,
in which we can replace the particle interactions by a set of simple
splitting kernels. Predictions beyond this simple approximation are an
active research field in view of the coming LHC
runs~\cite{Hartgring:2013jma,Li:2016yez,Hoche:2017hno,Dulat:2018vuy,
  Dasgupta:2018nvj,Dasgupta:2020fwr}. This progress motivates the
question what kind of fundamental QCD properties we can test in terms
of parton splittings defining relatively simple physics objects.

There exists a history of extracting QCD properties from
collider data. Before the LHC era, a global strategy combining
LEP-measurements like jet and event shapes, scaling violation,
fragmentation functions, $Z$-pole measurements, and $\tau$-decays with
low-energy $e^+ e^-$-data extracted the value of the strong coupling as
$\alpha_s(m_Z)=0.1211 \pm
0.0021$~\cite{Kluth:2003yz,Kluth:2006bw}. Obviously, this measurement
has since been improved by hadron collider and HERA data. In addition,
LEP data has been used for another fundamental QCD measurement, namely
QCD color factors or specifically quadratic $SU(3)$ Casimir invariants
as they appear in QCD splittings. They can be extracted from a variety
of jet or event shapes in 3-jet and 4-jet final states.  To start
with, measurements in 3-jet events by OPAL give $C_A/C_F = 2.232 \pm
0.14$~\cite{Abbiendi:2001us}, while the 3-jet measurements by DELPHI
lead to $C_A/C_F = 2.26 \pm 0.16$~\cite{Abreu:1999af}. Studies of the
electroweak 4-jet kinematic by ALEPH gives $C_A = 2.93 \pm 0.60$ and
$C_F = 1.35 \pm 0.27$~\cite{Heister:2002tq}, a similar analysis by
OPAL quotes $C_A = 3.02 \pm 0.56$ and $C_F = 1.34 \pm
0.30$~\cite{Abbiendi:2001qn} . In both 4-jet analyses there exists a
strong, positive correlation between $C_A$ and $C_F$.  Event
shapes~\cite{Brandt:1964sa,Farhi:1977sg,Parisi:1978eg,Donoghue:1979vi,Catani:1992jc,Dasgupta:2003iq},
similar to modern jet shapes~\cite{Feige:2012vc}, can be used to
extract the same parameters and give $C_A = 2.84 \pm 0.24$ and $C_F =
1.29 \pm 0.18$~\cite{Kluth:2000km}. The combined analysis
reports~\cite{Kluth:2003yz,Kluth:2006bw}
\begin{align}
  C_A = 2.89 \pm 0.21
  \quad \text{and} \quad 
  C_F = 1.30 \pm 0.09 \; ,
\label{eq:lep}
\end{align}
with both measurements being clearly systematics limited.

In this paper we propose a way to adapt these measurements for the LHC
era. An obvious path would be a comprehensive analysis of multi-jet
production, which would probably have to be combined with the global
extraction of parton densities together with the strong coupling
constant.  Instead of such a comprehensive global analysis we base our
study on parton shower data, which does not require us to understand
parton densities or mass effects or large electroweak corrections.
Our goal is to put the above LEP measurements into a context of QCD
measurements at the LHC and to develop a framework for learning the
properties of QCD splittings.  At the subjet level there exists a
range of observables for which we can compare precision predictions
with precision
measurements~\cite{Dasgupta:2013ihk,Adams:2015hiv,Marzani:2019hun}. On
the other hand, parton showers are a prime example for precision
simulations, so we will follow the orthogonal approach of extracting
fundamental QCD parameters using simulation-based inference.  The
inspiration for our analysis are new methods referred to as
likelihood-free inference at the event
level~\cite{Brehmer:2018kdj,Brehmer:2019xox}. In both cases, the
crucial ingredient is first-principle precision simulations with full
control over the underlying hypothesis and over its theoretical
self-consistency in describing the corresponding objects.

Technically, our BayesFlow approach~\cite{radev2020bayesflow} is based
on the conditional version~\cite{cinn,cinn2} of invertible networks
(INNs)~\cite{inn,coupling2,glow}, a specific realization of
normalizing
flows~\cite{nflow1,papamakarios2019normalizing,Kobyzev_2020,mller2018neural}. These
networks have been studied in relation to phase space
generation~\cite{Bothmann:2020ywa,Gao:2020vdv,Gao:2020zvv,Chen:2020nfb},
event generation~\cite{Verheyen:2020bjw}, anomaly
detection~\cite{Nachman:2020lpy}, detector and parton shower
unfolding~\cite{Bellagente:2020piv}, and density
estimation~\cite{Brehmer:2020vwc}. We will introduce our QCD inference
framework in Sec.~\ref{sec:inn} and illustrate our splitting kernel
measurements in Sec.~\ref{sec:toy}. Finally, we will attempt a more
realistic benchmarking for the \sherpa~\cite{Bothmann:2019yzt} shower
with hadronization and detector effects in Sec.~\ref{sec:det}.

\section{INN-Inference and BayesFlow}
\label{sec:inn}

\paragraph{INN}

The workhorse of our inference method is an invertible neural network
(INN) which realizes a normalizing flow~\cite{Kobyzev_2020} between
model parameters $m$ viewed as random vectors and a latent random
vector $z$.  Such an INN with the trainable parameters $\theta$
represents an easily invertible function $g_{\theta}(m)$ which
transforms $m$ into $z$, whereas its inverse $\bar{g}_{\theta}(z)$
transforms $z$ back into $m$. This way the INN simultaneously encodes
both directions of a bijective mapping between $m$ and $z$ via a
single set of parameter $\theta$ learned through gradient-based
optimization.

Coupling flows are a widely used invertible architecture, since they
are capable of learning highly expressive transformations with
tractable Jacobian determinants~\cite{Kobyzev_2020,glow}.  We
construct our INNs by composing multiple affine coupling
layers~\cite{coupling1,coupling2} into a composite invertible
architecture.  A single coupling layer $g_{\theta_j}$ splits its input
vector $m$ into two halves, $m = (m^A$, $m^B$), to obtain $z = (z^A$,
$z^B)$ via the bijective transformation
\begin{align}
\begin{pmatrix} z^A \\ z^B \end{pmatrix} = 
\begin{pmatrix}
m^A \odot e^{{s_2(m^B)}} + t_2(m^B) \\
m^B \odot e^{{s_1(z^A)}} + t_1(z^A)
\end{pmatrix}
\quad \Leftrightarrow \quad
\begin{pmatrix} m^A \\ m^B \end{pmatrix} = 
\begin{pmatrix}
(z^A - t_2(m^B)) \odot e^{{-s_2(m^B)}} \\
(z^B - t_1(z^A)) \odot e^{{-s_1(z^A)}} 
\end{pmatrix} \; .
\label{eq:layers}
\end{align}
By construction, this bijection works independently of the form of the
functions $s$ and $t$.  For our application, $s$ and $t$ are realized
via feed-forward neural networks with trainable parameters $\theta_j$
in each coupling layer.
The Jacobian of each coupling flow layer is the product of two triangular matrices
\begin{align}
\frac{\partial g_{\theta_j}(m)}{\partial m} =
\begin{pmatrix}
\one  & 0 \\
\text{finite} & \text{diag } e^{s_1(z^A)}
\end{pmatrix}
\begin{pmatrix}
\text{diag } e^{s_2(m^B)} & \text{finite} \\
0 & \one 
\end{pmatrix} \; ,
\label{eq:jacob}
\end{align}
%
%
%
%
making its determinant fast to compute.  Much effort has gone into
improving the efficiency of invertible coupling layers.  We use the
all-in-one coupling layer with three additional
features~\cite{coupling2,glow}.  First, each layer incorporates
a fixed permutation before splitting its input, to ensure that each
component in the final $z$ is influenced by each component of the
initial $m$.  Second, it includes a global affine transformation to
induce a bias and linear scaling.  Third, it applies a bijective soft
clamping after the exponential function in Eq.\eqref{eq:layers} to
prevent instabilities from divergent outputs~\cite{cinn}.

We combine multiple coupling layers to increase the expressiveness of
the learned transformation.  This is possible because a combination of
invertible functions is again invertible and its Jacobian is the
product of the individual Jacobians.  For $J$ coupling layers, our
composite INN is given by
\begin{align}
    z = g_{\theta_J} \circ g_{\theta_{J-1}} \circ \cdot\cdot\cdot \circ g_{\theta_1}(m)
\end{align}
with trainable parameters $\theta = (\theta_1,\dots,\theta_J)$ and the
inverse
\begin{align}
    m = \bar{g}_{\theta_1} \circ \cdot\cdot\cdot \circ \bar{g}_{\theta_{J-1}}\circ \bar{g}_{\theta_J}(z)
\end{align}
Such a composition can be viewed as transforming or normalizing a
complicated, intractable source distribution $P(m)$ into a much
simpler, tractable, $P(z)$ prescribed by the optimization criterion.

\paragraph{Conditional INN}

To recover model parameters from a set of measurements $x$ we need to
augment the INN architecture in two ways.  First, we turn the
invertible network into a conditional invertible network (cINN).  A
cINN still defines a bijective mapping between $m$ and $z$, but the
functions $s$ and $t$ in each coupling layer take a set of
measurements as an additional input, $z = g_{\theta}(m; x)$.  Second,
since the number of measurements can vary in practice, we introduce a
relatively small summary network $h_{\psi}$ with trainable parameters
$\psi$. It reduces measurements of variable size to fixed-size
vectors, $\tilde{x} = h_{\psi}(x)$, by respecting the probabilistic
symmetry of the measurements~\cite{radev2020bayesflow}.  For
independent measurements we use a permutation invariant summary
network such that its output is invariant under the ordering in
$x$~\cite{radev2020bayesflow}. The summary network does not have to be
invertible, since its output is concatenated with $m$ and fed to $s$
and $t$, but not directly mapped to $z$.  Moreover, the two networks
can be trained together to approximate the true parameter posterior
$P(m | x)$ via an approximate posterior $Q$ defined by the network
weights.  Due to the change of variable formula, this approximate
posterior is given by
\begin{align}
  Q(m | x) = P(z) \left|\det \left(\frac{\partial z}{\partial m}\right)\right|
  \quad \text{with} \quad z = g_{\theta}(m; h_{\psi}(x))
\label{eq:cinn}
\end{align} 
and it represents the probabilistic solution to the inverse inference problem.

Together, the cINN and the summary network minimize the expected
Kullback-Leibler divergence between the true and approximate
posterior.  Ignoring all terms that do not depend on the network
parameters, this corresponds to minimizing the expected negative
logarithm of the approximate posterior,
\begin{align}
\min_{\theta, \psi} \langle \mathbb{KL} \left( P(m | x)\,|| \,Q(m|x) \right) \rangle_{m,x} 
\sim \min_{\theta, \psi} \left\langle  -\log Q(m | x) \right\rangle_{m,x} + \text{const.}
\end{align}
Finally, we can apply a coordinate transformation for the bijective
mapping and enforce a Gaussian noise distribution with mean zero and
width one for the latent distribution $P(z)$, so the loss
function becomes
\begin{align}
  L(\theta, \psi) &= - \left\langle \log P(g_{\theta}(m; h_{\psi}(x)))
                + \log \left| \frac{\partial g_{\theta}(m;h_{\psi}(x))}{\partial m} \right|
     \right\rangle_{m,x} \notag \\
  &= - \left\langle - \frac{1}{2} \left\Vert g_\theta(m;h_{\psi}(x)) \right\Vert^2
      + \log \left| \frac{\partial g_{\theta}(m;h_{\psi}(x))}{\partial m} \right|
   \right\rangle_{m,x} \; . \label{eq:loss}
\end{align}
%
%
%
This loss guarantees that the networks recover the true posterior
under perfect convergence~\cite{radev2020bayesflow}.

\paragraph{Inference}

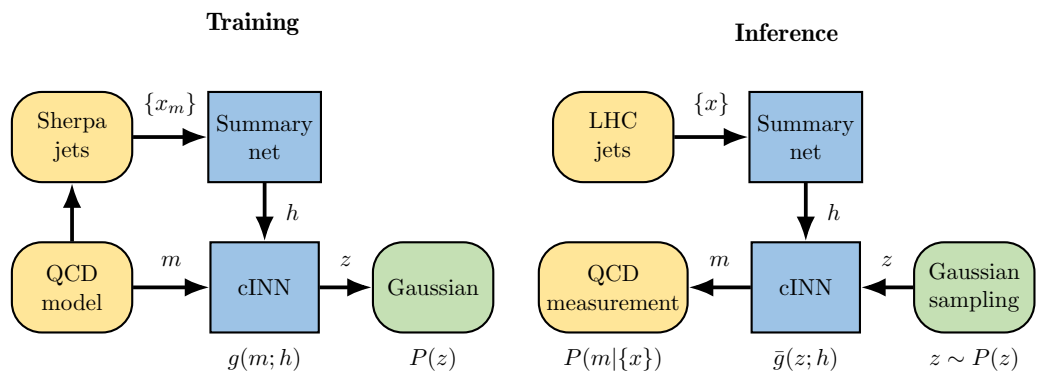
\begin{figure}[t]
\centering
\usetikzlibrary{arrows.meta,shapes}

\definecolor{Rcolor}{HTML}{E99595}
\definecolor{Gcolor}{HTML}{C5E0B4}
\definecolor{Bcolor}{HTML}{9DC3E6}
\definecolor{Ycolor}{HTML}{FFE699}

\tikzstyle{network} = [thick, rectangle, minimum width=1.8cm, minimum height=1.5cm, text centered, align=center, draw]
\tikzstyle{data} = [thick, rectangle, rounded corners=0.4cm, minimum width=2.0cm, minimum height=1.5cm, text centered, align=center,draw ]
\tikzstyle{arrow} = [thick,-{Latex[scale=1.0]}, line width=0.5mm]

\begin{tikzpicture}[node distance=2cm, scale=0.8, every node/.style={transform shape}]
\node (cinn) [network, fill=Bcolor] {cINN};
\node (summary) [network, above of = cinn, yshift=0.5cm, fill=Bcolor] {Summary\\net};
\node (sherpa) [data, left of = summary, xshift = -1.2cm, fill=Ycolor] {Sherpa\\jets};
\node (model) [data, left of = cinn, xshift = -1.2cm, fill=Ycolor] {QCD\\model};
\node (random) [data, right of = cinn, xshift = 0.8cm, fill=Gcolor] {Gaussian};

\draw [arrow, color=black] ([yshift=0em]summary.south) -- ([yshift=0em]cinn.north) node[midway,right,xshift=0.2cm]{$h$};
\draw [arrow, color=black] ([yshift=0em]sherpa.east) -- ([yshift=0em]summary.west) node[midway,above,yshift=0.2cm]{$\{x_m\}$};
\draw [arrow, color=black] ([yshift=0em]model.north) -- ([yshift=0em]sherpa.south);
\draw [arrow, color=black] ([yshift=0em]model.east) -- ([yshift=0em]cinn.west) node[midway,above,yshift=0.2cm]{$m$};
\draw [arrow, color=black] ([yshift=0em]cinn.east) -- ([yshift=0em]random.west) node[midway,above,yshift=0.2cm]{$z$};

\node [below of = cinn, yshift=0.8cm] {$g(m;h)$};
\node [below of = random, yshift=0.8cm] {$P(z)$};

\node[above,font=\large\bfseries, yshift = 0.7cm] at (current bounding box.north) {Training};

\end{tikzpicture}\nobreak\hspace{1.5em}%
\begin{tikzpicture}[node distance=2cm, scale=0.8, every node/.style={transform shape}]
\node (cinn) [network, fill=Bcolor] {cINN};
\node (summary) [network, above of = cinn, yshift=0.5cm, fill=Bcolor] {Summary\\net};
\node (jets) [data, left of = summary, xshift = -1.2cm, fill=Ycolor] {LHC\\jets};
\node (measurement) [data, left of = cinn, xshift = -1.2cm, fill=Ycolor] {QCD\\measurement};
\node (sampling) [data, right of = cinn, xshift = 0.8cm, fill=Gcolor] {Gaussian \\ sampling};
\node[above,font=\large\bfseries, yshift = 0.7cm] at (current bounding box.north) {Inference};

\draw [arrow, color=black] ([yshift=0em]summary.south) -- ([yshift=0em]cinn.north) node[midway,right,xshift=0.2cm]{$h$};
\draw [arrow, color=black] ([yshift=0em]jets.east) -- ([yshift=0em]summary.west) node[midway,above,yshift=0.2cm]{$\{x\}$};
\draw [arrow, color=black] ([yshift=0em]cinn.west) -- ([yshift=0em]measurement.east) node[midway,above,yshift=0.2cm]{$m$};
\draw [arrow, color=black] ([yshift=0em]sampling.west) -- ([yshift=0em]cinn.east) node[midway,above,yshift=0.2cm]{$z$};

\node [below of = measurement, yshift=0.8cm] {$P(m|\{x\})$};
\node [below of = cinn, yshift=0.8cm] {$\bar g(z;h)$};
\node [below of = random, yshift=0.8cm] {$z \sim P(z)$};
\end{tikzpicture}
\caption{BayesFlow setup of the cINN for training and
  inference~\cite{radev2020bayesflow}.}
\label{fig:network}
\end{figure}

BayesFlow~\cite{radev2020bayesflow} provides a cINN framework which we
can use to measure fundamental QCD parameters.  From the inversion of
a detector simulation and QCD radiation~\cite{Bellagente:2020piv} we
know how, given a single detector-level event, the cINN generates
samples from a probability distribution over the phase space of the
hard scattering.  For the jet inference presented in this paper, the
BayesFlow setup corresponds to this unfolding setup, in which we
replace the parton-level phase space with the model parameter space
and the detector-level phase space with the simulated data. In
Fig~\ref{fig:network} we give a graphical illustration of the
inference setup, for the training and the inference phases.

To train the BayesFlow networks we use the fact that we can
simulate an arbitrary number of jets fast. This allows us to employ
mini-batch gradient descent to approximate the expectation in the
above optimization criterion via its Monte-Carlo empirical mean.
Moreover, if we train the networks on jet samples of varying size, we
can use them on data samples with any size, as long as this size is
within the domain of the pre-defined distribution over sample sizes.
The networks will approximate the correct push-forward from a given
prior $P(m)$ in model space to a posterior $P(m | x)$ contingent on a
set of measurements $x$.  When the test sample size leaves the
training domain the posterior accuracy will degrade. In case we need to analyse larger data sets we can then follow the
Bayesian logic behind the BayesFlow framework~\cite{radev2020bayesflow} and use the 
posterior from an earlier measurement as a prior.

\section{Idealized jet measurements}
\label{sec:toy}

Before applying BayesFlow to LHC jets including hadronization and
detector simulation, we define our theory assumptions and test the
corresponding model on an idealized data set using a toy
shower~\cite{Hoche:2014rga}. That will give us an idea what kind of
measurement we could aim for and will also allow for some simple
benchmarking. We have checked that this toy shower agrees with the
full \sherpa shower, except that we do not include the effects from
the 2-loop cusp anomalous dimension.

\begin{table}[b!]
\centering
\begin{small} \begin{tabular}{l|c c}
\toprule
& Symbol & Value \\
\midrule
Number of parameters          & $L$ & $2, 3$ \\
Maximum number of constituents & $F$ & 13 \\
Jets per parameter point (variable/fixed)    & $M$ & $10^2~...~10^5$ / $10^4$\\
Batch size                    & $N$ & 16 \\
Batches per epoch             & $E$ & 6250 \\
Output dimension summary network     & $S$ & 32 \\
Fully connected summary net architecture & $S_i$ & 64,64,64,64,32,32 \\
Coupling layers & $n_\text{layers}$ & 5 \\
Fully connected coupling layer architecture & $s_i/t_i$ & 64,64,64 \\
Epochs & e  & $10~...~40$  \\
Decay steps (toy shower/PF flow) & $n_{s}$ & $200~...~500$ / $500~...~1000$\\
Learning rate after $t$ batches & $\eta_t$ & $10 ^{-3} \cdot 0.99^{\lfloor t/n_{s} \rfloor}$ \\
Training/testing points & & 100k / 10k  \\
\bottomrule
\end{tabular} \end{small}
\caption{Network setup and hyperparameters.}
\label{tab:para}
\end{table}

\paragraph{Theory setup}

The physics goal in our paper is to understand the QCD splittings
building up parton showers. In the leading collinear approximation
these kernels relate the amplitudes of an $n$-particle hard process
$\mat_n$ to the amplitude with an additional parton
$\mat_{n+1}$~\cite{Plehn:2015dqa}
\begin{align}
  \overline{\left|\mat_{n+1}\right|^2}
  \simeq \frac{2 g_s^2}{p_a^2}
   \; \hat{P}(z,y) \; \overline{\left|\mat_n\right|^2} \; ,
\label{eq:defkernel}
\end{align}
where $g_s$ is the strong coupling, $p_a^2 = (p + k)^2$ the invariant
mass of the splitting parton, and $\hat{P}(z,y)$ the un-regularized
splitting kernel. It depends on the energy fraction $z$ and the
momentum transfer $y$, which in combination with a Catani-Seymour
spectator momentum $p_s$ can be combined to the transverse momentum in
the splitting,
\begin{align}
        z &=\frac{p p_s}{p p_s+k p_s} \notag \\
        y &=\frac{p k}{p k + p p_s+ kp_s}
        \quad \Rightarrow \quad yz(1-z) \propto p_T^2 \; .
\end{align}
In massless QCD some of the kernels $\hat{P}$ include infrared
divergences. They can be partially fractioned to remove soft
double counting,
giving us the three QCD splittings~\cite{Catani:1996jh}
\begin{align}
P_{qq}(z,y) &= C_F \left[ D_{qq} \frac{2z(1-y)}{1-z(1-y)} + F_{qq} (1-z) + C_{qq} yz(1-z) \right] \notag \\
P_{gg}(z,y) &= 2 C_A \bigg[ D_{gg} \left(\frac{z(1-y)}{1-z(1-y)} + \frac{(1-z)(1-y)}{1-(1-z)(1-y)}\right) 
          + F_{gg} z(1-z) + C_{gg} yz(1-z) \bigg] \notag \\
 P_{gq}(z,y) &= T_R \left[ F_{qq} \left(z^2 + (1-z)^2 \right) + C_{gq} yz(1-z) \right] \; .
\label{eq:qcd_kernels}
\end{align}
In this form we include a set of parameters which to leading order in
perturbative QCD are given by
\begin{align}\label{eq:def_params}
  D_{qq,gg} = 1
  \qqquad
  F_{qq,gg} = 1
  \qqquad
  C_{qq,gg,gq} = 0 \; .
\end{align}
The splitting kernels given in Eq.\eqref{eq:qcd_kernels} define the
fundamental physics hypothesis of our measurements, which should
generalize the $C_A/C_F$ studies from
LEP~\cite{Kluth:2003yz,Kluth:2006bw}. This hypothesis is flexible
enough to accomodate precision predictions consistently with the
kinematics of parton shower data at the LHC. Concerning its
uniqueness, in standard parton showers, $D$ is typically modified to
include a universal $K$-factor that coincides with the two-loop cusp
anomalous dimension and resums sub-leading logarithms arising from the
collinear splitting of soft gluons~\cite{Catani:1990rr}. For
simplicity, we will set these terms to zero in our toy shower.  Within
\sherpa, they are included through a modified running coupling. The
second term reflects the leading terms in $p_T$, in our case truncated
in the strong coupling. The rest terms $C_{ij}$ are, in our case,
defined by the appearance of $p_T^2$.  Generally, we only consider
Eq.\eqref{eq:qcd_kernels} as a first attempt for an appropriate
theory hypothesis, which might have to be slightly modified according
to the precision simulation framework used for the actual
analysis. Another motivations for a modified theory hypothesis could
be specific parametrizations to, for instance, incorporate quantum
effects or $1 \to 3$ splittings. We skip this option because we will
see that already the global rest terms of Eq.\eqref{eq:qcd_kernels}
challenge our simulated data.  As alluded to in the Introduction, a
caveat concerning the pre-defined theory hypothesis is common to all
simulation-based or likelihood-free analyses.

We will vary the parameters in Eq.\eqref{eq:def_params}
away from the leading order QCD prediction, always making sure
that the splitting kernels give positive splitting probabilities all
over the collinear phase space by setting negative kernel values to zero. Given that the numerically leading
contribution comes from the regularized pole, we can approximately
identify the measurement of $D_{qq}$ and $D_{gg}$ with measurements of
$C_F$ and $C_A$, as quoted in Eq.\eqref{eq:lep}.

\paragraph{Data and network}

To understand the proposed measurement in a controlled setup we
simulate the on-shell process
\begin{align}
  e^{+} e^{-} \to Z \to q \bar{q}
\end{align}
assuming massless quarks and combined with a fast approximate parton
shower cutoff at 1~GeV. Its phase space is completely defined by the
scattering angle. For each event we apply the parton shower to one of
the outgoing quarks, such that the second quark acts as the spectator
for the the first splitting and we only consider one jet. For 
our jets sample we generally have
\begin{align}
  p_{T,j} < \frac{m_Z}{2} \; ,
\end{align}
with the majority of jets at the upper boundary.  After that, any
other parton can act as the spectator. For this simple setup a jet
reconstruction is not necessary, since we only simulate a single
shower, and we neither include hadronization nor detector effects.

The network then analyses the set of outgoing momenta except for the
initial spectator momentum.  The list of constituents includes up
to $F$ entries, and is zero-padded or cropped. For our training data
we scan the parameter space $\{ D_{ij},F_{ij},C_{ij} \}$ with $L=2$
and 3~dimensions. For each parameter point we generate $M$
probabilistic showers.  To observe the correct posterior contraction
with the size of the test sample we train the network with variable
$M$. During the training we use batches of size $N$. The input to the
summary network per batch are $N \times M \times F$ 4-vectors. The output of
the summary networks is mean-pooled over $M$ and has dimension $S$ for
each batch, plus the value of $\sqrt{M}$, 
so $(S+1)$ entries per batch,  if the posterior contraction is trained.

\begin{figure}[t]
  \includegraphics[width=0.495\textwidth]{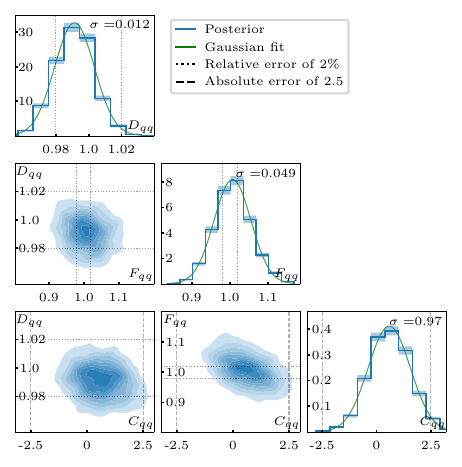}
  \includegraphics[width=0.495\textwidth]{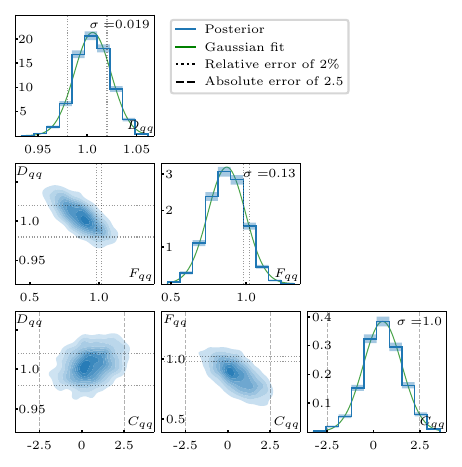}
  \caption{Posterior probabilities for the toy shower, gluon radiation
    only, $\{ D_{qq}, F_{qq},C_{qq} \}$. We assume SM-like jets and
    show results for truth-sorting (left) and for $k_T$-sorting
    (right).}
  \label{fig:dqq_fqq_cqq}
\end{figure}

The distribution of the number of jets $M$ over the $N$ batches can be
adapted to the problem. We find that distributing the batches with
$1/M$ is effective to counter the computational effort at high $M$. We
will explicitly show that we retain enough high-$M$ information to
guarantee the correct scaling of the error.

The cINN then provides a bijective mapping of the $L$-dimensional
parameter space to the latent space of the same dimension, again per
batch. The latent space is forced into a Gaussian noise form, so we
can sample from it to compute the probability distribution for a given
set of $M_\text{eval}$ showers in model space.
Values $M_{eval}$ not included in the training will lead to unstable
results, if $\sqrt{M}$ was added to the summary network output. All
parameters of the network architecture and the hyperparameters are
given in Tab.~\ref{tab:para}. For the cINN we combine five coupling
layers. The internal networks of the coupling layers, $s_{1/2}$ and
$t_{1/2}$, are three fully connected layers with ELU activation. The
summary network is built out of six fully connected layers with ReLU
activation, ELU activation in the last layer, followed by average
pooling. We use the Adam optimizer~\cite{adam} with an exponentially
decaying learning rate.

\paragraph{Sorting}

Even though this constitutes an information backdoor, we first study
what the network can extract if we order the constituents following
their appearance in the shower. We refer to this unrealistic ordering
of the constituents as truth-sorting. This means that after a
splitting the daughter constituents are either appended to the end of
the list or replace the mother momentum. Which daughter momentum
overwrites the mother momentum is chosen by the showering
algorithm. To avoid this backdoor we construct a similar ordering from
the shower history given by a
$k_T$-algorithm~\cite{Dokshitzer:1997in,fastjet}, referred to as
$k_T$-sorting. Since we simulate only a single jet, no jet radius
has to be specified. We start with the first splitting and follow the hard
constituents as the particles with the highest energy fraction in each
splitting to determine the first entry of the list. The next entry is generated 
by following the hard constituents originating from the softer constituent of
the earliest splitting. This is done for every splitting going from first to last. If an appearing parton is already assigned to the list, it is not 
assigned again.

\paragraph{Gluon-radiation shower}

For our first test we restrict the shower to the $P_{qq}$ kernel of
Eq.\eqref{eq:qcd_kernels}, implying that a hard quark successively
radiates collinear and soft gluons. This way our 3-dimensional model
space is given by
\begin{align}
  \{ D_{qq}, F_{qq},C_{qq} \} \; .
\end{align}
For the prior in model space we start with a uniform distribution over
$[0.5, 2] \times [0, 4] \times [-10,10]$ and train the network for
100000 randomly distributed points in model space.

\begin{figure}[t]
  \includegraphics[width=0.32\textwidth]{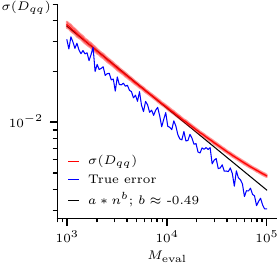}
  \includegraphics[width=0.32\textwidth]{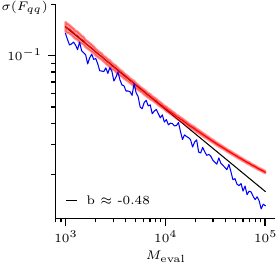}
  \includegraphics[width=0.32\textwidth]{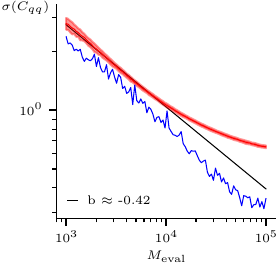} \\
  \includegraphics[width=0.32\textwidth]{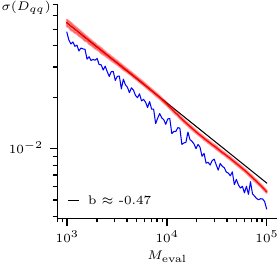}
  \includegraphics[width=0.32\textwidth]{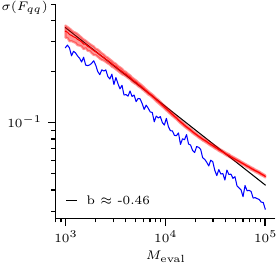}
  \includegraphics[width=0.32\textwidth]{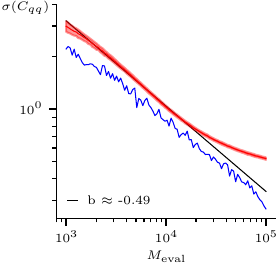}
  \caption{Uncertainty on $\{ D_{qq}, F_{qq}, C_{qq}\}$ for gluon
    radiation only, as a function of the number of test jets. We show
    the standard deviation of the posterior (red) and the the absolute
    difference between the estimated and true parameters (blue) for
    truth-sorting (upper) and for $k_T$-sorting (lower). The black
    line is a fit to the posterior.}
  \label{fig:scaling}
\end{figure}

In Fig.~\ref{fig:dqq_fqq_cqq} we show the distribution of the
posterior probabilities for $10^2 ... 10^5$ training jets per
parameter point and 10000 test jets assuming SM-values. The fit
confirms that all 1-dimensional posteriors are approximately
Gaussian. Correlations among them are weak. As expected, we are more
sensitive for the truth-sorting with its information backdoor. The
reduced performance with the $k_T$-sorting indicates that
pre-processing of the data plays an important role, and that
information from jet algorithms should help. For the $k_T$-sorting the
best-measured parameter in our toy model is the regularized divergence
with a Gaussian standard deviation $\sigma(D_{qq}) =
0.019$.\footnote{All numerical results in this paper are also
  collected in Tab.~\ref{tab:error}.} The finite terms are slightly
harder to extract with $\sigma(F_{qq}) = 0.13$. Finally, the rest term
with its assumed $p_T$-suppression comes with an even larger error,
$\sigma(C_{qq}) = 1.0$. These increasing errors reflect the
hierarchical structure of the splitting kernel. In addition to the
reduced performance we also see that the information lost between
truth-sorting and $k_T$-sorting induces visible correlations between
the extracted model parameters. This correlation explains some of the
loss in performance for instance in the $D_{qq}$ vs $F_{qq}$ plane,
where the widths of the 1-dimensional posterior distributions are
driven by the integration over the other parameters.

In Fig.~\ref{fig:scaling} we show how the errors on these three
model parameter change with the statistics of the test data set. Given
the size of the training samples, $M = 10^2~...~10^5$, we evaluate
$M_\text{eval} = 10^3~...~10^5$ SM-like jets and find that the
Gaussian errors $\sigma(D_{qq})$, $\sigma (F_{qq})$, and
$\sigma (C_{qq})$ all scale like $1/\sqrt{M_\text{eval}}$. This is
expected for a statistically limited measurement. We also check the
consistency of the network by comparing the reported standard
deviation with the deviation between the central estimates and the
truth. Altogether, the network performs exactly as expected, with the
exception of a slight degradation in the challenging rest term
$C_{qq}$ towards large test statistics.

\paragraph{QCD splittings}

\begin{figure}[t]
  \includegraphics[width=0.495\textwidth]{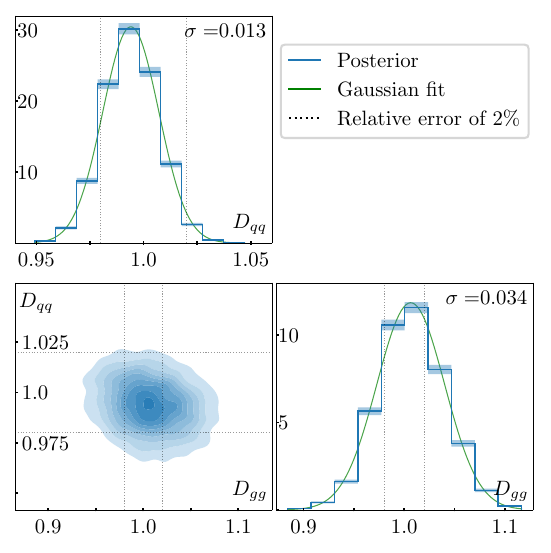}
  \includegraphics[width=0.495\textwidth]{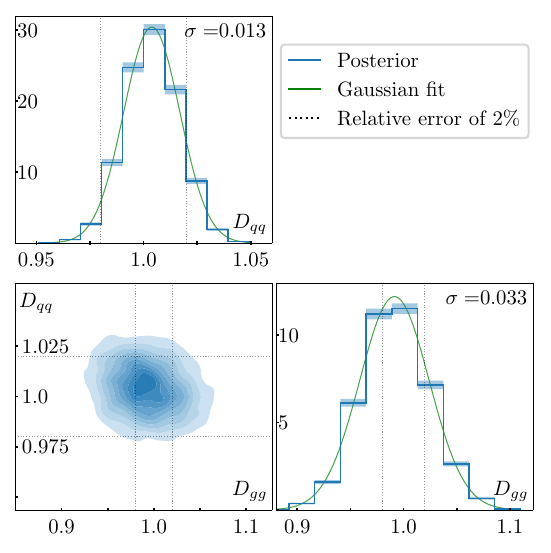}
  \caption{Posterior probabilities for the toy shower, soft-collinear
    leading terms for all QCD splittings, $\{ D_{qq}, D_{gg}\}$. We
    assume SM-like jets and show results for truth-sorting (left) and
    for $k_T$-sorting (right).}
  \label{fig:dqq_dgg}
\end{figure}

In a second step, we include all three QCD splitting kernels from
Eq.\eqref{eq:qcd_kernels} and extract the soft-collinear divergences,
\begin{align}
  \{ D_{qq}, D_{gg} \} \; .
\end{align}
Assuming that the leading logarithms really dominate the splittings
and the sub-jet features, this measurement corresponds to measuring
the two Casimirs $C_F$ and $C_A$. Because we only consider quark-jets
from $Z$-decays, we expect the measurement of $D_{qq}$ or $C_F$ to be
better, which is also what we observe in
Fig.~\ref{fig:dqq_dgg}. Notably, for truth-sorting and for
$k_T$-sorting there is no correlation between the two measurements,
unlike for the standard LEP results.

\begin{figure}[t]
  \includegraphics[width=0.495\textwidth]{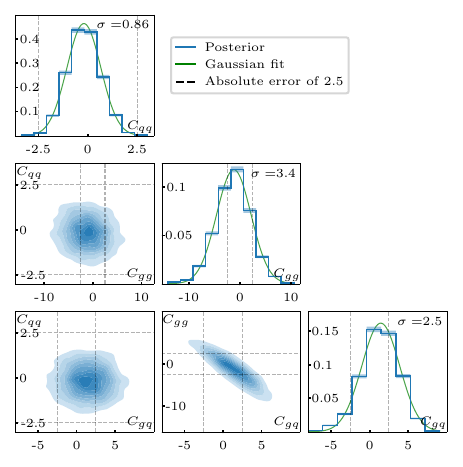}
  \includegraphics[width=0.495\textwidth]{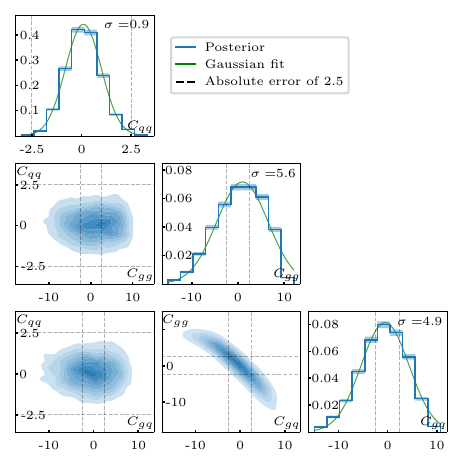}
  \caption{Posterior probabilities for the toy shower,
    $p_T$-suppressed rest terms for all QCD splittings, $\{ C_{qq},
    C_{gg}, C_{gq} \}$.  We assume SM-like jets and show results for
    truth-sorting (left) and for $k_T$-sorting (right).}
  \label{fig:cqq_cgg_cgq}
\end{figure}

For the final test on our toy model we include all QCD splitting
kernels from Eq.\eqref{eq:qcd_kernels} and determine the three
$p_T$-suppressed rest terms
\begin{align}
  \{ C_{qq},C_{qg},C_{gg} \} \; .
\end{align}
Hypothesis-wise this means that we assume that our predictions for the
two leading contributions hold, and we want to estimate the size of an
unknown contribution at higher power in $p_T$. The network
is the same as for the gluon-radiation shower, with the results shown
in Fig.~\ref{fig:cqq_cgg_cgq}.  For $C_{qq}$ we first see that in the
absence of the dominant contribution, the error drops slightly
with respect to Fig.~\ref{fig:dqq_fqq_cqq}. The reason is that it is
challenging for the network to disentangle the hierarchical structure
of $\{ D_{qq},F_{qq},C_{qq} \}$ in Fig.~\ref{fig:dqq_fqq_cqq}.  For
the other two rest terms, $C_{gg}$ and $C_{gq}$, we find significantly
larger 1-dimensional errors and, related to these a strong
anti-correlation. This correlation already exists for the
truth-sorting case, so we expect it to remain in any realistic
measurement.

\paragraph{High-level observables}

\begin{figure}[t]
  \includegraphics[width=0.495\textwidth]{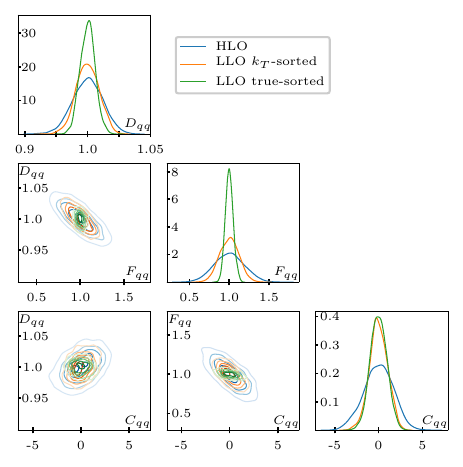}
  \includegraphics[width=0.495\textwidth]{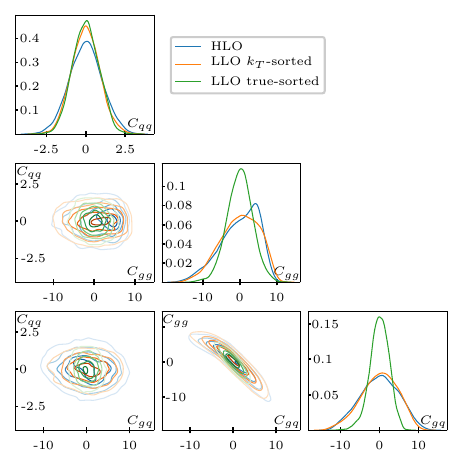}
  \caption{Posterior probabilities from the low-level observables with
    two sortings and the high-level observables given in
    Eqs.\eqref{eq:qg_obs1} and \eqref{eq:qg_obs2}. In the left panel
    we assume gluon radiation only, $\{ D_{qq}, F_{qq},C_{qq} \}$,
    corresponding to Fig.~\ref{fig:dqq_fqq_cqq}. In the right panel we
    measure the $p_T$-suppressed terms in all QCD splittings, $\{
    C_{qq}, C_{gg}, C_{gq} \}$, corresponding to
    Fig.~\ref{fig:cqq_cgg_cgq}.}
  \label{fig:HLOvsLLO}
\end{figure}

To judge the impact of the low-level network input we can use the same
setup as before, but feed high-level observables into the summary
network. We use a set of six such
observables~\cite{Kasieczka:2018lwf}, not all of them infrared and
collinear safe.  The simplest high-level observable for subjet analysis
tracks the size of the splitting probabilities in terms of particle
multiplicity ($n_\text{PF}$)~\cite{Frye:2017yrw}. The width of the
distributed radiation or girth is denoted at
$w_\text{PF}$~\cite{Gallicchio:2010dq}. The effect of the soft
divergence can be measured using $p_TD$~\cite{CMS:2013kfa}.  In
addition, the two-point energy correlator $C_{0.2}$ is designed to
separate quarks and gluons with an optimized power of $\Delta
R_{ij}$~\cite{Larkoski:2013eya}.  This defines a set of four standard
observables
\begin{align}
n_\text{PF} &=  \sum_i 1 \qqquad &
w_\text{PF} &= \frac{\sum_i p_{T,i} \Delta R_{i,\text{jet}}}{ \sum_i p_{T,i}}  
\notag \\
p_TD &= \frac{\sqrt{\sum_i p_{T,i}^2 }}{ \sum_i p_{T,i} } \qqquad &
C_{0.2} &= \frac{\sum_{ij} E_{T,i} E_{T,j} (\Delta R_{ij})^{0.2} }{\sum_i E_{T,i}^2} \; .
\label{eq:qg_obs1}
\end{align}
In addition, we evaluate the highest fraction of $p_{T,\text{jet}}$
contained in a single jet constituent and the minimum number of
constituents which contain 95\% of
$p_{T,\text{jet}}$~\cite{Pumplin:1991kc},
\begin{align}
x_\text{max} 
\qqquad \text{and} \qqquad 
N_{95} \; .
\label{eq:qg_obs2}
\end{align}
The latter is obviously correlated with the number of constituents
$n_\text{PF}$.

In the left panels of Fig.~\ref{fig:HLOvsLLO} we compare the
posteriors from the low-level observables and the six high-level
observables for the gluon-radiation shower. The low-level results
correspond to Fig.~\ref{fig:dqq_fqq_cqq}, and we remind ourselves that
the truth-sorting with its information backdoor clearly leads to the
best results. On the other hand, the $k_T$-sorting still delivers much
better results than the high-level observables. In particular, the
additional information from the complete low-level information passed
through the summary net reduces the correlations between the three
measured parameters.

The right panels of Fig.~\ref{fig:HLOvsLLO} show the same
$p_T$-suppressed rest terms as we see in Fig.~\ref{fig:cqq_cgg_cgq},
but including the projected measurements from the high-level
observables. Again, the truth-sorting should not be taken as a
realistic benchmark, but even the $k_T$-sorting avoids the
non-Gaussian structures we see for the high-level observables. Aside
from that, the more democratic structure of the parameter set $\{
C_{qq}, C_{gg}, C_{gq} \}$ implies that the high-level and
low-level observables show more similar performance.

This comparison between the high-level and the low-level observables
should be taken with a grain of salt. First, we know that
pre-processing plays a role for the low-level network input, and one
could hope to recover a performance closer to the
truth-ordering. Second, the non-Gaussian posterior of $C_{gg}$ from
the high-level observables suggests that not all trainings might be as
stable as the successful training we show here.

\section{Hadronization and detector}
\label{sec:det}

\begin{figure}[b!]
  \centering
  \includegraphics[width=0.4\textwidth]{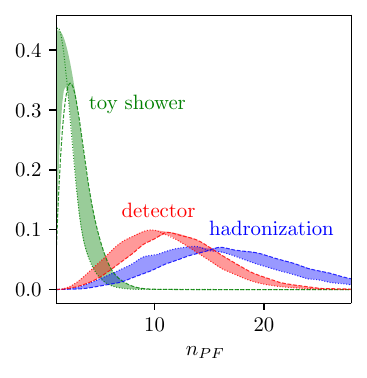}
  \hspace*{0.05\textwidth}
  \includegraphics[width=0.4\textwidth]{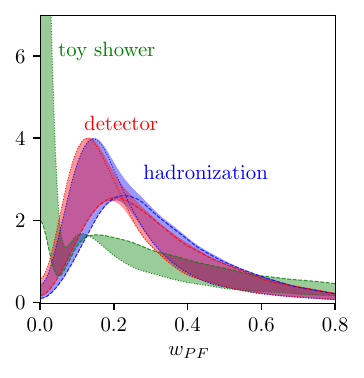} \\
  \includegraphics[width=0.4\textwidth]{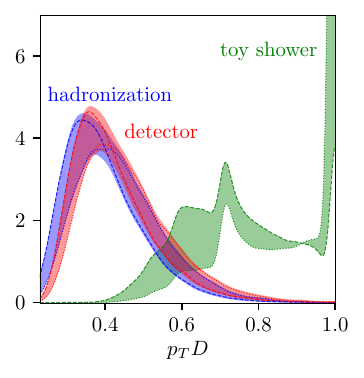}
  \hspace*{0.05\textwidth}
  \includegraphics[width=0.4\textwidth]{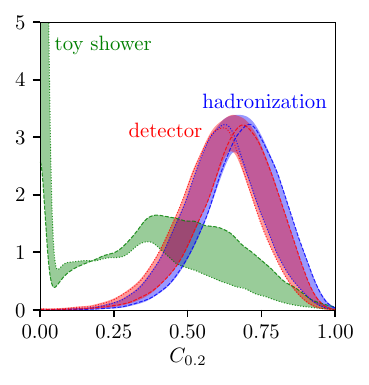} 
  \caption{High-level observables $n_{PF}$, $w_{PF}$, $p_TD$, and
    $C_{0.2}$ for 100k jets. We show results for the toy shower, the
    \sherpa shower with hadronization, and including detector effects
    with \delphes.  Bands show the variation of $D_{qq} = 0.5.~...~2$
    (dotted and dashed).}
  \label{fig:generator_comp}
\end{figure}

Obviously, the results from the quark-induced toy shower are not what
we can expect from an LHC analysis. Already for the comparison with
the LEP measurements we need to include hadronization rather than
cutting off the QCD splittings at a fixed scale of 1~GeV. In addition,
we know that the LHC detectors cannot compete with the $e^+ e^-$
environment, but on the other hand the available number of jets will
eventually be much larger. Given the promising results for the toy
shower the question is how well the analysis would work in a more
realistic environment.

For a more realistic simulation we turn to a modified version of
\sherpa2.2.10~\cite{Bothmann:2019yzt} and again generate the process
\begin{align}
  e^+ e^- \rightarrow q \bar q
  \qquad \text{with} \quad q=u,d,s \; ,
\end{align}
without the weakly decaying heavy quarks.  The leptonic initial state
plays no role for our jet analysis and allows us to ignore initial
state radiation.  The parton shower is modified to include our
parameterized splitting functions and has a cutoff at 1~GeV.  Within
\sherpa we still use the modified splitting kernels of
Eq.\eqref{eq:qcd_kernels} and vary different parameter sets while
setting all the others to their SM-values. Unlike for the toy shower
we do not remove QCD splittings for the \sherpa case.  Without a
detector simulation we save the 4-momenta of hadrons, photons and
charged leptons.  The maximum number of constituents, jets per
parameter points etc are identical to our toy shower analysis, and our
$k_T$-sorting algorithm is applied to these 4-momenta.  In a second
step we include LHC detector effects using
\delphes3.4.2~\cite{deFavereau:2013fsa} with the default ATLAS
card. Now we save the 4-momenta of all particle flow (PF) objects. The jets
are constructed with \fastjet3.3.4~\cite{Cacciari:2011ma}, either
processing the hadronization output or the \delphes output as $R=1.2$
anti-$k_T$ jets, giving the spectrum
\begin{align}
  p_{T,j} = 20~\gev~...~\frac{m_Z}{2} \; ,
\end{align}
We select only one jet per event for the jet sample. By LHC standards
these jets are soft, and since we are testing the structure of QCD
splittings, harder jets would include much more information. Because
this additional information will at some point be balanced by
challenging the calorimeter resolution we stick to this probably
over-conservative setup. We also ignore underlying event and pile-up,
because standard tools are going to be far from an optimal working
point for subjet analyses using low-level observables.

\begin{figure}[t]
  \includegraphics[width=0.495\textwidth]{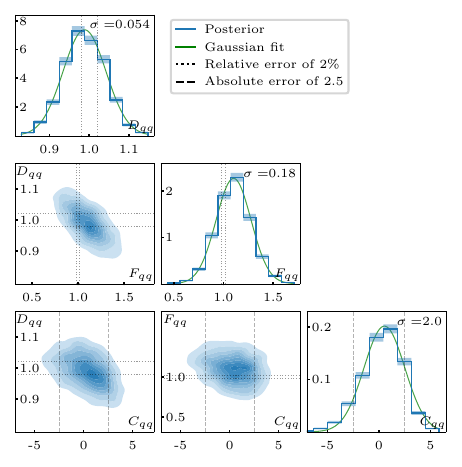}
  \includegraphics[width=0.495\textwidth]{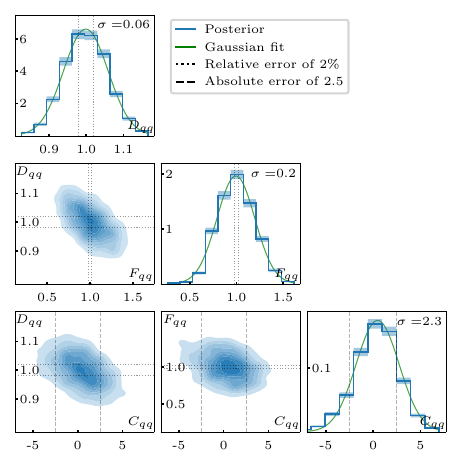}
  \caption{Posterior probabilities for the \sherpa shower, varying the
    gluon radiation parameters only, $\{ D_{qq}, F_{qq}, C_{qq}
    \}$. We assume SM-like jets and show results without \delphes
    detector simulation (left) and including detector effects
    (right).}
  \label{fig:sherpa_dqq_dqq_cqq}
\end{figure}

To illustrate the physics behind our proposed measurement, we show the
high-level observables from Eq.\eqref{eq:qg_obs1} for the toy shower,
after hadronization, and after detector effects in
Fig.~\ref{fig:generator_comp}. The bands are defined by a variation
$D_{qq} = 0.5.~...~2$, to illustrate the dependence on the splitting
kernels. The number of constituents $n_\text{PF}$ generally increases
with $D_{qq}$.  The toy shower does not generate a very large number
of splittings. Hadronization increases the number of constituents
significantly, but this effect has nothing to do with QCD
splittings. The detector simulation with its resolution and thresholds
again leads to a slight decrease. The width of the constituent
distribution, $w_\text{PF}$, is small for the toy shower, with a peak
once the toy shower generates enough splittings. An increase in
$D_{qq}$ moves the distribution away from very small values.
Hadronization enhances the peak around $w_\text{PF} \approx 0.2$,
driven by the hadron decays, and the detector effects have a limited
effect because of the explicit $p_T$-weighting.  For $p_TD$ a single
hard object gives $p_TD = 1$ and adding a soft constituent leads to a
downward shift. The small number of QCD splittings leads to a second
peak structure around $p_TD \sim 0.7$ for the toy shower, but the
entire toy-level distribution has to be taken with a grain of
salt. Hadronization then induces the typical shape with a broad
maximum below 0.5, again with little impact from the detector effects.
Finally, the constituent-constituent correlation $C_{0.2}$ loses all
toy-level events at small values when we include hadronization, and
the broad feature around $C_{0.2} \sim 0.4$ becomes more narrow and
moves to values around 0.6. As a side remark, this variable is
particularly effective to distinguish jets from hard quarks and hard
gluons, because the two peak structures are relatively well separated
with gluons giving larger values of $C_{0.2}$.

\begin{figure}[t]
    \includegraphics[width=0.32\textwidth]{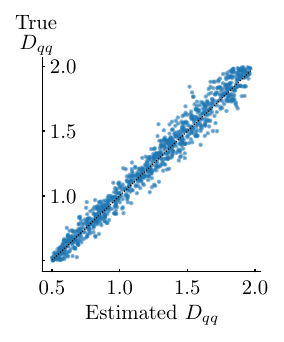}
    \includegraphics[width=0.32\textwidth]{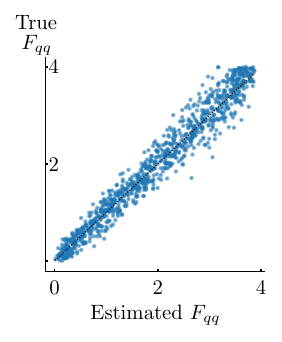}
    \includegraphics[width=0.32\textwidth]{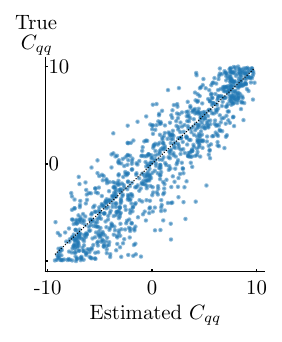}
    \caption{Posterior means vs true values of $D_{qq}$ (left),
      $F_{qq}$ (middle) and $C_{qq}$ (right) for a test data set with
      1000 parameters sets drawn from the prior with 10000 events each
      calculated including detector simulations in analogy to
      Fig.~\ref{fig:sherpa_dqq_dqq_cqq}.}
    \label{fig:sherpa_delphes_dqq_fqq_dqq_est_vs_true}
\end{figure}

The main message from Fig.~\ref{fig:generator_comp} is that from a QCD
point of view the hadronization effects are qualitatively and
quantitatively far more important than the detector
effects. Therefore, we split our study into two parts. First, we shift
from the toy shower to the full \sherpa
shower~\cite{Bothmann:2019yzt}, including hadronization. Next, we add
detector effects using \delphes~\cite{deFavereau:2013fsa} with the
default ATLAS card. Unlike for the toy shower, we now vary the
parameters for gluon radiation,
\begin{align}
  \{ D_{qq}, F_{qq},C_{qq} \}
  \qquad \text{(gluon radiation varied)} \; ,
\end{align}
while keeping the other splittings fixed to their SM-values.  The
results are shown in Fig.~\ref{fig:sherpa_dqq_dqq_cqq}. We apply our
$k_T$-sorting throughout this section, to ensure that there is no
information backdoor. Compared to the toy-shower in
Fig.~\ref{fig:dqq_fqq_cqq} all error bars for the \sherpa shower are
increased by a factor two to four. Intriguingly, the detector
resolution adds very little to the uncertainties in the case of our
relatively soft jets.

Throughout our analysis we have always assumed that testing our
network on SM-like jets is representative of the whole parameter range
the model is trained on. For this three-parameter case with a variable
quark-gluon splitting we also evaluate 1000 measurements over the
whole range covered by the prior. For each parameter point we generate
$M=10^4$ showers, sample 2000 points from the latent space to the
measurement, and identify the actual measurement with the average over
these 2000 measurements. In
Fig.~\ref{fig:sherpa_delphes_dqq_fqq_dqq_est_vs_true} we correlate the
true values with the measured values and find that they track each
other without a bias, but with a spread corresponding to the known
error bars.

\begin{figure}[t]
  \includegraphics[width=0.495\textwidth]{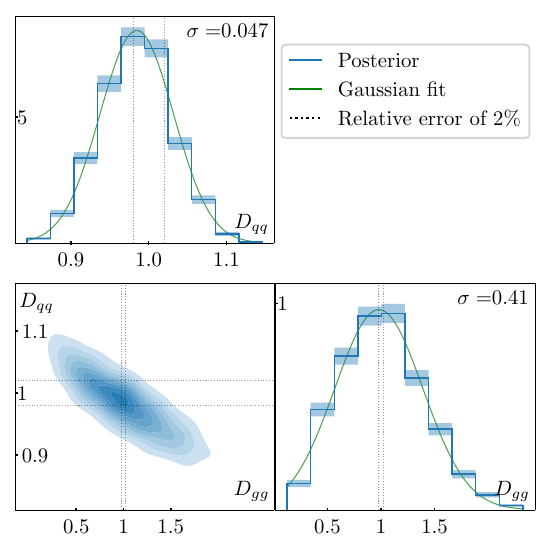}
  \includegraphics[width=0.495\textwidth]{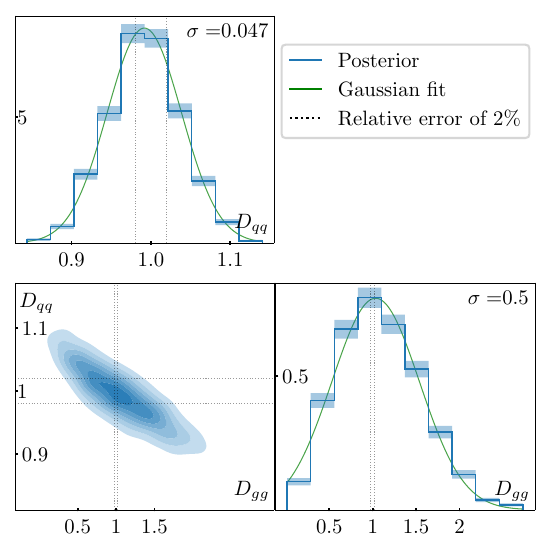}
  \caption{Posterior probabilities for the \sherpa shower,
    soft-collinear leading terms for all QCD splittings, $\{ D_{qq},
    D_{gg}\}$. We assume SM-like jets and show results without
    \delphes detector simulation (left) and including detector effects
    (right).}
  \label{fig:sherpa_dqq_dgg}
\end{figure}

Next, we allow for all QCD splittings included in the \sherpa shower
and measure the leading soft-collinear contributions, corresponding to
measuring $C_F$ and $C_A$ from a sample of quark-induced showers. The
combined measurement of the varied parameters
\begin{align}
  \{ D_{qq}, D_{gg} \} 
  \qquad \text{(soft-collinear varied)} 
\end{align}
is shown in Fig.~\ref{fig:sherpa_dqq_dgg} and can be directly compared
to the toy shower results from Fig.~\ref{fig:dqq_dgg}. Here we see a
significant degradation of the \sherpa measurements, especially in
$D_{gg}$. This is at least partly due to the correlation between the
two extracted model parameters which we do not observe for the toy
setup. This correlation is, if anything, slightly enhanced by the
detector effects, but as before the effect of the hadronization
clearly dominates. For an actual LHC measurement the correlation could
be easily removed by combining a quark-dominated and a gluon-dominated
jet sample. This is why we also report the measurement for one free
model parameter at a time in Tab.~\ref{tab:error}, indicating that at
the LHC we might be able to measure the leading contributions to the
splitting kernels in Eq.\eqref{eq:qcd_kernels} to the few per-cent
level.

\begin{figure}[t]
  \includegraphics[width=0.495\textwidth]{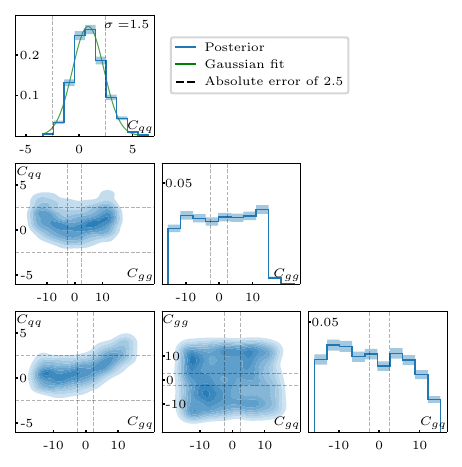}
  \includegraphics[width=0.495\textwidth]{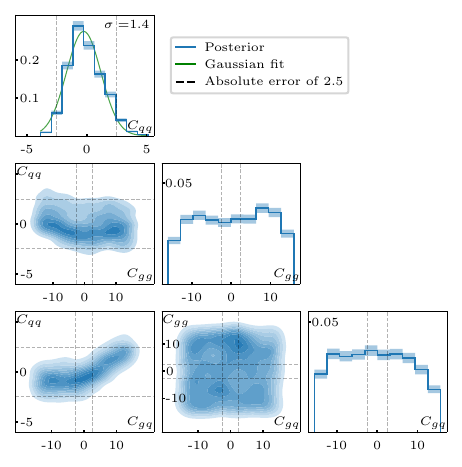}
  \caption{Posterior probabilities for the \sherpa shower,
    $p_T$-suppressed rest terms for all QCD splittings, $\{ C_{qq},
    C_{gg}, C_{gq} \}$.  We assume SM-like jets and show results
    without \delphes detector simulation (left) and including detector
    effects (right).}
  \label{fig:sherpa_cqq_cgg_cgq}
\end{figure}

Our final hypothesis is that we know the leading and constant terms in
all QCD splitting functions, but want to measure possible
deviations. This corresponds to varying and then extracting the
explicitly $p_T$-suppressed parameters
\begin{align}
  \{ C_{qq},C_{qg},C_{gg} \} 
  \qquad \text{(rest terms varied)} 
\end{align}
from the full \sherpa shower. Unlike for the other parameters, the
rest terms are defined around zero, with an explicit
$p_T$-suppression. This way we can argue that we do not expect any of
the $C_{ij}$ to be significantly larger than one. If that should be
the case, our expansion is not correct and the perturbative QCD
description of the respective jet sample has a problem.  Just looking
at $\sigma(C_{qq})$, where the error is roughly a factor two
larger than for the toy shower in Fig.~\ref{fig:cqq_cgg_cgq}, this
task is clearly in reach, even for our conservative setup. The two
other rest terms are essentially invisible.  Some of this can likely
be cured by combining a quark-dominated and a gluon-dominated sample,
where the latter will provide a measurement of $C_{gg}$.\bigskip

\begin{table}[b!]
\centering
\begin{small} \begin{tabular}{lc| l ll ll l |l l l l}
\toprule
\multicolumn{2}{c|}{Setup \& Parameter} & \multicolumn{6}{c|}{Toy shower} & \multicolumn{4}{c}{\sherpa} \\
& & \multicolumn{2}{l}{Truth-sorted} & \multicolumn{2}{l}{$k_T$-sorted} & \multicolumn{2}{l|}{HLO} & \multicolumn{2}{l}{Hadronized} & \multicolumn{2}{l}{Detector-level} \\
\midrule
\multirow{2}{*}{$\{D_{qq},F_{qq},C_{qq} \}$} &
  $\sigma(D_{qq})$ & $0.012$ & ($0.013$)& $0.019$ & ($0.013$) & $0.024$ & ($0.015$)& $0.054$ & ($0.025$) & $0.060$ & ($0.03$)\\
 &$\sigma(F_{qq})$ & $0.05$ & ($0.04$)& $0.16$ & ($0.07$)& $0.19$ & ($0.08$) & $0.18$ & ($0.09$)& $0.20$ & ($0.1$)\\
 &$\sigma(C_{qq})$ & $0.97$ & ($0.8$)& $1.04$ & ($0.8$)& $1.7$ & ($1.0$)& $2.0$ & ($1.2$)& $2.3$ & ($1.4$)\\[2mm]
\multirow{2}{*}{$\{ D_{qq}, D_{gg} \}$} &
  $\sigma(D_{qq})$ & $0.013$ & ($0.013$)& $0.013$ & ($0.013$)& $0.013$ & ($0.013$) & $0.047$ & ($0.025$)& $0.047$ & ($0.025$)\\
 &$\sigma(D_{gg})$ & $0.034$ & ($0.034$)& $0.033$ & ($0.033$)& $0.035$ & ($0.035$)& $0.41$ & ($0.23$) & $0.50$ & ($0.25$)\\[2mm]
\multirow{2}{*}{$\{ C_{qq}, C_{gg}, C_{gq} \}$} &
  $\sigma(C_{qq})$ & $0.86$ & ($0.8$)& $0.90$ & ($0.8$)& $1.0$ & ($1.0$)& $1.5$ & ($1.0$)& $1.4$ & ($0.9$)\\
 &$\sigma(C_{gg})$ & $3.4$ & ($1.4$)& $5.6$ & ($1.7$)& $5.4*$ & ($1.7$)& $*$ & $*$ & $*$ & $*$ \\
 &$\sigma(C_{gq})$ & $2.7$ & ($1.1$) & $4.9$ & ($1.4$)& $5.2*$ & ($1.4$)& $*$ & $*$ & $*$ & $*$ \\
\bottomrule
\end{tabular} \end{small}
\caption{Error on the extracted QCD splitting kernels from 10k events
  in the different setups: gluon radiation only, soft-collinear
  leading contributions, and $p_T$-suppressed rest terms.  The
  truth-sorting includes an information backdoor through the ordering
  of the inputs. The asterisk denotes a non-Gaussian posterior. The
  error in parentheses assumes one variable splitting parameter at a
  time.}
\label{tab:error}
\end{table}

In Tab.~\ref{tab:error} we collect all numerical results from this
paper. We test three hypotheses, (i) all terms in the gluon radiation
off a quark, (ii) the leading terms of the quark and gluon splitting,
corresponding for instance to the color Casimirs $C_F$ and $C_A$, and (iii)
the $p_T$-suppressed rest terms of the three splitting kernels.  The
results for the toy shower indicate that the information on the QCD
splitting kernels is indeed included in the low-level
observables. Obviously, it is easier to extract the leading,
regularized pole terms than the finite terms. The rest terms are
expected to be zero for the perturbative prediction, so we expect them
to be at most of order one. One of the interesting results from the
toy shower is that the multi-dimensional posteriors show hardly any
correlations, except for the two rest terms $C_{gq}$ and
$C_{gg}$. This correlation could be cured by adding gluon-dominated
showers in our analysis. Moving to the \sherpa shower we first notice
that hadronization has a much more degrading effect than detector
effects. While it is still possible to determine the splitting
function for gluon emission off a quark and the regularized
soft-collinear divergences, we do not have enough sensitivity to
constrain all three rest terms. However, $C_{qq}$ is within reach, and
$C_{gg}$ should be testable if we include a gluon-dominated jet
sample.

\section{Outlook}

The gold standard of LHC physics is our ability to understand all
aspects of the recorded events in terms of fundamental physics. Parton
showers, or parton splittings, are part of every LHC analysis. In
spite of an active subjet physics program and in spite of significant
theoretical progress, we do not have a systematic set of measurements
of their simple underlying QCD predictions, even though similar
analyses based on jet and event shapes do exist from LEP.

In this paper we have proposed a systematic approach to measuring QCD
splittings, including an appropriate technique based on modern machine
learning. To define a viable and consistent theory hypothesis, we have
parameterized the known splitting kernels into the leading,
logarithmically enhanced term, the finite term known in perturbative
QCD, and a rest term. Traditionally, the leading term could be
identified with the measurement of QCD color factors. The finite term
reflects the simple description of parton splittings as a Markov
process, while the rest term would allow us to parameterize for
instance quantum effects or higher splittings. Expanding the theory
hypothesis accordingly would be a natural step in refining any
simulation-based analysis.

We have then shown that for a toy shower, modelled after \sherpa, we
can measure all these contributions from low-level observables of a
jet sample. For a realistic version we saw that hadronization has the
biggest effect on our measurement, bigger than the expected detector
effects for relatively soft jets. The challenge will be to extract the
rest terms beyond the standard QCD predictions, to test the quality of
the perturbative QCD prediction.

Our analysis method is based on machine learning, specifically an
invertible network conditioned on a small summary network. After
training, we can use the invertible network to sample the model
parameter space and construct a posterior probability based on a set
of jets. While we study SM-like jets throughout our analysis, the
network produces the correct posterior for all jets covered by the
original parameter scan.

Our analysis is not meant to be the final word on ML-measurements of
fundamental QCD properties from LHC jets.  Natural next steps, aside
from testing our methodology on actual data would be a second,
gluon-initiated jet sample and an additional harder jet sample. While
the former will get rid of the remaining correlations in the model
parameters, the latter should allow us to optimize the interplay of
the energy range covered by the shower and the calorimeter resolution.
Our current setup is also not efficient in analyzing millions of jets,
because unlike standard likelihood methods it does not scale well with
additional data. This is the downside of directly extracting the
posterior distribution.

\begin{center} \textbf{Acknowledgments} \end{center}

We would like to thank Uli Uwer for fun discussions in general, on the
superiority of LHCb, and on the corresponding LEP measurements.  The
research of AB and TP is supported by the Deutsche
Forschungsgemeinschaft (DFG, German Research Foundation) under grant
396021762 -- TRR~257 \textsl{Particle Physics Phenomenology after the
  Higgs Discovery}.
The authors acknowledge support by the state of Baden-Württemberg through bwHPC
and the German Research Foundation (DFG) through grant no INST 39/963-1 FUGG (bwForCluster NEMO).
This research was supported by the Fermi National Accelerator Laboratory
(Fermilab), a U.S. Department of Energy, Office of Science, HEP User Facility.
Fermilab is managed by Fermi Research Alliance, LLC (FRA), acting
under Contract No. DE--AC02--07CH11359.

\bibliography{literature}

\providecommand{\href}[2]{#2}\begingroup\raggedright\begin{thebibliography}{10}

\bibitem{Lonnblad:1990bi}
L.~Lonnblad, C.~Peterson, and T.~Rognvaldsson, {\it {Finding Gluon Jets With a
  Neural Trigger}},  \href{http://dx.doi.org/10.1103/PhysRevLett.65.1321}{Phys.
  Rev. Lett. {\bfseries 65} (1990)  1321}.

\bibitem{Csabai:1990tg}
I.~Csabai, F.~Czako, and Z.~Fodor, {\it {Quark and gluon jet separation using
  neural networks}},  \href{http://dx.doi.org/10.1103/PhysRevD.44.R1905}{Phys.
  Rev. D {\bfseries 44} (1991)  1905}.

\bibitem{Cogan:2014oua}
J.~Cogan, M.~Kagan, E.~Strauss, and A.~Schwarztman, {\it {Jet-Images: Computer
  Vision Inspired Techniques for Jet Tagging}},
  \href{http://dx.doi.org/10.1007/JHEP02(2015)118}{JHEP {\bfseries 02} (2015)
  118}, \href{http://arxiv.org/abs/1407.5675}{{arXiv:1407.5675 [hep-ph]}}.

\bibitem{deOliveira:2015xxd}
L.~de~Oliveira, M.~Kagan, L.~Mackey, B.~Nachman, and A.~Schwartzman, {\it
  {Jet-images --- deep learning edition}},
  \href{http://dx.doi.org/10.1007/JHEP07(2016)069}{JHEP {\bfseries 07} (2016)
  069}, \href{http://arxiv.org/abs/1511.05190}{{arXiv:1511.05190 [hep-ph]}}.

\bibitem{Baldi:2016fql}
P.~Baldi, K.~Bauer, C.~Eng, P.~Sadowski, and D.~Whiteson, {\it {Jet
  Substructure Classification in High-Energy Physics with Deep Neural
  Networks}},  \href{http://dx.doi.org/10.1103/PhysRevD.93.094034}{Phys. Rev. D
  {\bfseries 93} (2016) 9, 094034},
  \href{http://arxiv.org/abs/1603.09349}{{arXiv:1603.09349 [hep-ex]}}.

\bibitem{Komiske:2016rsd}
P.~T. Komiske, E.~M. Metodiev, and M.~D. Schwartz, {\it {Deep learning in
  color: towards automated quark/gluon jet discrimination}},
  \href{http://dx.doi.org/10.1007/JHEP01(2017)110}{JHEP {\bfseries 01} (2017)
  110}, \href{http://arxiv.org/abs/1612.01551}{{arXiv:1612.01551 [hep-ph]}}.

\bibitem{Kasieczka:2017nvn}
G.~Kasieczka, T.~Plehn, M.~Russell, and T.~Schell, {\it {Deep-learning Top
  Taggers or The End of QCD?}},
  \href{http://dx.doi.org/10.1007/JHEP05(2017)006}{JHEP {\bfseries 05} (2017)
  006}, \href{http://arxiv.org/abs/1701.08784}{{arXiv:1701.08784 [hep-ph]}}.

\bibitem{Macaluso:2018tck}
S.~Macaluso and D.~Shih, {\it {Pulling Out All the Tops with Computer Vision
  and Deep Learning}},  \href{http://dx.doi.org/10.1007/JHEP10(2018)121}{JHEP
  {\bfseries 10} (2018)  121},
  \href{http://arxiv.org/abs/1803.00107}{{arXiv:1803.00107 [hep-ph]}}.

\bibitem{Almeida:2015jua}
L.~G. Almeida, M.~Backovi\'c, M.~Cliche, S.~J. Lee, and M.~Perelstein, {\it
  {Playing Tag with ANN: Boosted Top Identification with Pattern Recognition}},
   \href{http://dx.doi.org/10.1007/JHEP07(2015)086}{JHEP {\bfseries 07} (2015)
  086}, \href{http://arxiv.org/abs/1501.05968}{{arXiv:1501.05968 [hep-ph]}}.

\bibitem{Butter:2017cot}
A.~Butter, G.~Kasieczka, T.~Plehn, and M.~Russell, {\it {Deep-learned Top
  Tagging with a Lorentz Layer}},
  \href{http://dx.doi.org/10.21468/SciPostPhys.5.3.028}{SciPost Phys.
  {\bfseries 5} (2018) 3, 028},
  \href{http://arxiv.org/abs/1707.08966}{{arXiv:1707.08966 [hep-ph]}}.

\bibitem{Pearkes:2017hku}
J.~Pearkes, W.~Fedorko, A.~Lister, and C.~Gay, {\it {Jet Constituents for Deep
  Neural Network Based Top Quark Tagging}},
  \href{http://arxiv.org/abs/1704.02124}{{arXiv:1704.02124 [hep-ex]}}.

\bibitem{Erdmann:2018shi}
M.~Erdmann, E.~Geiser, Y.~Rath, and M.~Rieger, {\it {Lorentz Boost Networks:
  Autonomous Physics-Inspired Feature Engineering}},
  \href{http://dx.doi.org/10.1088/1748-0221/14/06/P06006}{JINST {\bfseries 14}
  (2019) 06, P06006}, \href{http://arxiv.org/abs/1812.09722}{{arXiv:1812.09722
  [hep-ex]}}.

\bibitem{Louppe:2017ipp}
G.~Louppe, K.~Cho, C.~Becot, and K.~Cranmer, {\it {QCD-Aware Recursive Neural
  Networks for Jet Physics}},
  \href{http://dx.doi.org/10.1007/JHEP01(2019)057}{JHEP {\bfseries 01} (2019)
  057}, \href{http://arxiv.org/abs/1702.00748}{{arXiv:1702.00748 [hep-ph]}}.

\bibitem{Komiske:2018cqr}
P.~T. Komiske, E.~M. Metodiev, and J.~Thaler, {\it {Energy Flow Networks: Deep
  Sets for Particle Jets}},
  \href{http://dx.doi.org/10.1007/JHEP01(2019)121}{JHEP {\bfseries 01} (2019)
  121},
\href{http://arxiv.org/abs/1810.05165}{{arXiv:1810.05165 [hep-ph]}}.

\bibitem{Qu:2019gqs}
H.~Qu and L.~Gouskos, {\it {ParticleNet: Jet Tagging via Particle Clouds}},
  \href{http://dx.doi.org/10.1103/PhysRevD.101.056019}{Phys. Rev. D {\bfseries
  101} (2020) 5, 056019},
  \href{http://arxiv.org/abs/1902.08570}{{arXiv:1902.08570 [hep-ph]}}.

\bibitem{Kasieczka:2019dbj}
A.~Butter {\em et al.}, {\it {The Machine Learning Landscape of Top Taggers}},
  \href{http://dx.doi.org/10.21468/SciPostPhys.7.1.014}{SciPost Phys.
  {\bfseries 7} (2019)  014},
  \href{http://arxiv.org/abs/1902.09914}{{arXiv:1902.09914 [hep-ph]}}.

\bibitem{Bollweg:2019skg}
S.~Bollweg, M.~Haussmann, G.~Kasieczka, M.~Luchmann, T.~Plehn, and J.~Thompson,
  {\it {Deep-Learning Jets with Uncertainties and More}},
  \href{http://dx.doi.org/10.21468/SciPostPhys.8.1.006}{SciPost Phys.
  {\bfseries 8} (2020) 1, 006},
  \href{http://arxiv.org/abs/1904.10004}{{arXiv:1904.10004 [hep-ph]}}.

\bibitem{Kasieczka:2020vlh}
G.~Kasieczka, M.~Luchmann, F.~Otterpohl, and T.~Plehn, {\it {Per-Object
  Systematics using Deep-Learned Calibration}},
  \href{http://arxiv.org/abs/2003.11099}{{arXiv:2003.11099 [hep-ph]}}.

\bibitem{Dreyer:2018nbf}
F.~A. Dreyer, G.~P. Salam, and G.~Soyez, {\it {The Lund Jet Plane}},
  \href{http://dx.doi.org/10.1007/JHEP12(2018)064}{JHEP {\bfseries 12} (2018)
  064}, \href{http://arxiv.org/abs/1807.04758}{{arXiv:1807.04758 [hep-ph]}}.

\bibitem{Lai:2020byl}
Y.~S. Lai, D.~Neill, M.~P\l{}osko\'n, and F.~Ringer, {\it {Explainable machine
  learning of the underlying physics of high-energy particle collisions}},
  \href{http://arxiv.org/abs/2012.06582}{{arXiv:2012.06582 [hep-ph]}}.

\bibitem{Marchesini:1987cf}
G.~Marchesini and B.~Webber, {\it {Monte Carlo Simulation of General Hard
  Processes with Coherent QCD Radiation}},
  \href{http://dx.doi.org/10.1016/0550-3213(88)90089-2}{Nucl. Phys. B
  {\bfseries 310} (1988)  461}.

\bibitem{Hartgring:2013jma}
L.~Hartgring, E.~Laenen, and P.~Skands, {\it {Antenna Showers with One-Loop
  Matrix Elements}},  \href{http://dx.doi.org/10.1007/JHEP10(2013)127}{JHEP
  {\bfseries 10} (2013)  127},
  \href{http://arxiv.org/abs/1303.4974}{{arXiv:1303.4974 [hep-ph]}}.

\bibitem{Li:2016yez}
H.~T. Li and P.~Skands, {\it {A framework for second-order parton showers}},
  \href{http://dx.doi.org/10.1016/j.physletb.2017.05.011}{Phys. Lett. B
  {\bfseries 771} (2017)  59},
  \href{http://arxiv.org/abs/1611.00013}{{arXiv:1611.00013 [hep-ph]}}.

\bibitem{Hoche:2017hno}
S.~H\"oche, F.~Krauss, and S.~Prestel, {\it {Implementing NLO DGLAP evolution
  in Parton Showers}},  \href{http://dx.doi.org/10.1007/JHEP10(2017)093}{JHEP
  {\bfseries 10} (2017)  093},
  \href{http://arxiv.org/abs/1705.00982}{{arXiv:1705.00982 [hep-ph]}}.

\bibitem{Dulat:2018vuy}
F.~Dulat, S.~H\"oche, and S.~Prestel, {\it {Leading-Color Fully Differential
  Two-Loop Soft Corrections to QCD Dipole Showers}},
  \href{http://dx.doi.org/10.1103/PhysRevD.98.074013}{Phys. Rev. D {\bfseries
  98} (2018) 7, 074013},
  \href{http://arxiv.org/abs/1805.03757}{{arXiv:1805.03757 [hep-ph]}}.

\bibitem{Dasgupta:2018nvj}
M.~Dasgupta, F.~A. Dreyer, K.~Hamilton, P.~F. Monni, and G.~P. Salam, {\it
  {Logarithmic accuracy of parton showers: a fixed-order study}},
  \href{http://dx.doi.org/10.1007/JHEP09(2018)033}{JHEP {\bfseries 09} (2018)
  033}, \href{http://arxiv.org/abs/1805.09327}{{arXiv:1805.09327 [hep-ph]}}.
  [Erratum: JHEP 03, 083 (2020)].

\bibitem{Dasgupta:2020fwr}
M.~Dasgupta, F.~A. Dreyer, K.~Hamilton, P.~F. Monni, G.~P. Salam, and G.~Soyez,
  {\it {Parton showers beyond leading logarithmic accuracy}},
  \href{http://dx.doi.org/10.1103/PhysRevLett.125.052002}{Phys. Rev. Lett.
  {\bfseries 125} (2020) 5, 052002},
  \href{http://arxiv.org/abs/2002.11114}{{arXiv:2002.11114 [hep-ph]}}.

\bibitem{Kluth:2003yz}
S.~Kluth, {\it {Jet physics in e+ e- annihilation from 14-GeV to 209-GeV}},
  \href{http://dx.doi.org/10.1016/j.nuclphysbps.2004.04.134}{Nucl. Phys. B
  Proc. Suppl. {\bfseries 133} (2004)  36},
  \href{http://arxiv.org/abs/hep-ex/0309070}{{arXiv:hep-ex/0309070}}.

\bibitem{Kluth:2006bw}
S.~Kluth, {\it {Tests of Quantum Chromo Dynamics at e+ e- Colliders}},
  \href{http://dx.doi.org/10.1088/0034-4885/69/6/R04}{Rept. Prog. Phys.
  {\bfseries 69} (2006)  1771},
  \href{http://arxiv.org/abs/hep-ex/0603011}{{arXiv:hep-ex/0603011}}.

\bibitem{Abbiendi:2001us}
OPAL, G.~Abbiendi {\em et al.}, {\it {Particle multiplicity of unbiased gluon
  jets from $e^{+} e^{-}$ three jet events}},
  \href{http://dx.doi.org/10.1007/s100520200926}{Eur. Phys. J. C {\bfseries 23}
  (2002)  597},
  \href{http://arxiv.org/abs/hep-ex/0111013}{{arXiv:hep-ex/0111013}}.

\bibitem{Abreu:1999af}
DELPHI, P.~Abreu {\em et al.}, {\it {Measurement of the gluon fragmentation
  function and a comparison of the scaling violation in gluon and quark jets}},
   \href{http://dx.doi.org/10.1007/s100520050719}{Eur. Phys. J. C {\bfseries
  13} (2000)  573}.

\bibitem{Heister:2002tq}
ALEPH, A.~Heister {\em et al.}, {\it {Measurements of the strong coupling
  constant and the QCD color factors using four jet observables from hadronic Z
  decays}},  \href{http://dx.doi.org/10.1140/epjc/s2002-01114-2}{Eur. Phys. J.
  C {\bfseries 27} (2003)  1}.

\bibitem{Abbiendi:2001qn}
OPAL, G.~Abbiendi {\em et al.}, {\it {A Simultaneous measurement of the QCD
  color factors and the strong coupling}},
  \href{http://dx.doi.org/10.1007/s100520100699}{Eur. Phys. J. C {\bfseries 20}
  (2001)  601},
  \href{http://arxiv.org/abs/hep-ex/0101044}{{arXiv:hep-ex/0101044}}.

\bibitem{Brandt:1964sa}
S.~Brandt, C.~Peyrou, R.~Sosnowski, and A.~Wroblewski, {\it {The Principal axis
  of jets. An Attempt to analyze high-energy collisions as two-body
  processes}},  \href{http://dx.doi.org/10.1016/0031-9163(64)91176-X}{Phys.
  Lett. {\bfseries 12} (1964)  57}.

\bibitem{Farhi:1977sg}
E.~Farhi, {\it {A QCD Test for Jets}},
  \href{http://dx.doi.org/10.1103/PhysRevLett.39.1587}{Phys. Rev. Lett.
  {\bfseries 39} (1977)  1587}.

\bibitem{Parisi:1978eg}
G.~Parisi, {\it {Super Inclusive Cross-Sections}},
  \href{http://dx.doi.org/10.1016/0370-2693(78)90061-8}{Phys. Lett. B
  {\bfseries 74} (1978)  65}.

\bibitem{Donoghue:1979vi}
J.~F. Donoghue, F.~Low, and S.-Y. Pi, {\it {Tensor Analysis of Hadronic Jets in
  Quantum Chromodynamics}},
  \href{http://dx.doi.org/10.1103/PhysRevD.20.2759}{Phys. Rev. D {\bfseries 20}
  (1979)  2759}.

\bibitem{Catani:1992jc}
S.~Catani, G.~Turnock, and B.~Webber, {\it {Jet broadening measures in $e^{+}
  e^{-}$ annihilation}},
  \href{http://dx.doi.org/10.1016/0370-2693(92)91565-Q}{Phys. Lett. B
  {\bfseries 295} (1992)  269}.

\bibitem{Dasgupta:2003iq}
M.~Dasgupta and G.~P. Salam, {\it {Event shapes in e+ e- annihilation and deep
  inelastic scattering}},
  \href{http://dx.doi.org/10.1088/0954-3899/30/5/R01}{J. Phys. G {\bfseries 30}
  (2004)  R143},
  \href{http://arxiv.org/abs/hep-ph/0312283}{{arXiv:hep-ph/0312283}}.

\bibitem{Feige:2012vc}
I.~Feige, M.~D. Schwartz, I.~W. Stewart, and J.~Thaler, {\it {Precision Jet
  Substructure from Boosted Event Shapes}},
  \href{http://dx.doi.org/10.1103/PhysRevLett.109.092001}{Phys. Rev. Lett.
  {\bfseries 109} (2012)  092001},
  \href{http://arxiv.org/abs/1204.3898}{{arXiv:1204.3898 [hep-ph]}}.

\bibitem{Kluth:2000km}
S.~Kluth, P.~Movilla~Fernandez, S.~Bethke, C.~Pahl, and P.~Pfeifenschneider,
  {\it {A Measurement of the QCD color factors using event shape distributions
  at s**(1/2) = 14-GeV to 189-GeV}},
  \href{http://dx.doi.org/10.1007/s100520100742}{Eur. Phys. J. C {\bfseries 21}
  (2001)  199},
  \href{http://arxiv.org/abs/hep-ex/0012044}{{arXiv:hep-ex/0012044}}.

\bibitem{Dasgupta:2013ihk}
M.~Dasgupta, A.~Fregoso, S.~Marzani, and G.~P. Salam, {\it {Towards an
  understanding of jet substructure}},
  \href{http://dx.doi.org/10.1007/JHEP09(2013)029}{JHEP {\bfseries 09} (2013)
  029},
\href{http://arxiv.org/abs/1307.0007}{{arXiv:1307.0007 [hep-ph]}}.

\bibitem{Adams:2015hiv}
D.~Adams {\em et al.}, {\it {Towards an Understanding of the Correlations in
  Jet Substructure}},
  \href{http://dx.doi.org/10.1140/epjc/s10052-015-3587-2}{Eur. Phys. J. C
  {\bfseries 75} (2015) 9, 409},
  \href{http://arxiv.org/abs/1504.00679}{{arXiv:1504.00679 [hep-ph]}}.

\bibitem{Marzani:2019hun}
S.~Marzani, G.~Soyez, and M.~Spannowsky,
  \href{http://dx.doi.org/10.1007/978-3-030-15709-8}{{\em {Looking inside jets:
  an introduction to jet substructure and boosted-object phenomenology}}},
  vol.~958.
\newblock Springer, 2019.
\newblock \href{http://arxiv.org/abs/1901.10342}{{arXiv:1901.10342 [hep-ph]}}.

\bibitem{Brehmer:2018kdj}
J.~Brehmer, K.~Cranmer, G.~Louppe, and J.~Pavez, {\it {Constraining Effective
  Field Theories with Machine Learning}},
  \href{http://dx.doi.org/10.1103/PhysRevLett.121.111801}{Phys. Rev. Lett.
  {\bfseries 121} (2018) 11, 111801},
  \href{http://arxiv.org/abs/1805.00013}{{arXiv:1805.00013 [hep-ph]}}.

\bibitem{Brehmer:2019xox}
J.~Brehmer, F.~Kling, I.~Espejo, and K.~Cranmer, {\it {MadMiner: Machine
  learning-based inference for particle physics}},
  \href{http://dx.doi.org/10.1007/s41781-020-0035-2}{Comput. Softw. Big Sci.
  {\bfseries 4} (2020) 1, 3},
  \href{http://arxiv.org/abs/1907.10621}{{arXiv:1907.10621 [hep-ph]}}.

\bibitem{radev2020bayesflow}
S.~T. Radev, U.~K. Mertens, A.~Voss, L.~Ardizzone, and U.~Köthe, {\it
  Bayesflow: Learning complex stochastic models with invertible neural
  networks},  \href{http://arxiv.org/abs/2003.06281}{{arXiv:2003.06281
  [stat.ML]}}.

\bibitem{cinn}
L.~Ardizzone, C.~Lüth, J.~Kruse, C.~Rother, and U.~Köthe, {\it Guided image
  generation with conditional invertible neural networks},
  \href{http://arxiv.org/abs/1907.02392}{{arXiv:1907.02392 [cs.CV]}}.

\bibitem{cinn2}
C.~Winkler, D.~Worrall, E.~Hoogeboom, and M.~Welling, {\it Learning likelihoods
  with conditional normalizing flows},
  \href{http://arxiv.org/abs/1912.00042}{{arXiv:1912.00042 [cs.LG]}}.

\bibitem{inn}
L.~Ardizzone, J.~Kruse, S.~Wirkert, D.~Rahner, E.~W. Pellegrini, R.~S. Klessen,
  L.~Maier-Hein, C.~Rother, and U.~Köthe, {\it Analyzing inverse problems with
  invertible neural networks},
  \href{http://arxiv.org/abs/1808.04730}{{arXiv:1808.04730 [cs.LG]}}.

\bibitem{coupling2}
L.~Dinh, J.~Sohl-Dickstein, and S.~Bengio, {\it Density estimation using real
  nvp},  \href{http://arxiv.org/abs/1605.08803}{{arXiv:1605.08803 [cs.LG]}}.

\bibitem{glow}
D.~P. Kingma and P.~Dhariwal, {\it Glow: Generative flow with invertible 1x1
  convolutions},  \href{http://arxiv.org/abs/1807.03039}{{arXiv:1807.03039
  [stat.ML]}}.

\bibitem{nflow1}
D.~J. Rezende and S.~Mohamed, {\it Variational inference with normalizing
  flows},  \href{http://arxiv.org/abs/1505.05770}{{arXiv:1505.05770
  [stat.ML]}}.

\bibitem{papamakarios2019normalizing}
G.~Papamakarios, E.~Nalisnick, D.~J. Rezende, S.~Mohamed, and
  B.~Lakshminarayanan, {\it Normalizing flows for probabilistic modeling and
  inference},  \href{http://arxiv.org/abs/1912.02762}{{arXiv:1912.02762
  [stat.ML]}}.

\bibitem{Kobyzev_2020}
I.~Kobyzev, S.~Prince, and M.~Brubaker, {\it Normalizing flows: An introduction
  and review of current methods},
  \href{http://dx.doi.org/10.1109/TPAMI.2020.2992934}{\href{http://dx.doi.org/10.1109/tpami.2020.2992934}{IEEE
  Transactions on Pattern Analysis and Machine Intelligence (2020)  1–1}}.

\bibitem{mller2018neural}
T.~Müller, B.~McWilliams, F.~Rousselle, M.~Gross, and J.~Novák, {\it Neural
  importance sampling},
  \href{http://arxiv.org/abs/1808.03856}{{arXiv:1808.03856 [cs.LG]}}.

\bibitem{Bothmann:2020ywa}
E.~Bothmann, T.~Janssen, M.~Knobbe, T.~Schmale, and S.~Schumann, {\it
  {Exploring phase space with Neural Importance Sampling}},
  \href{http://dx.doi.org/10.21468/SciPostPhys.8.4.069}{SciPost Phys.
  {\bfseries 8} (2020) 4, 069},
  \href{http://arxiv.org/abs/2001.05478}{{arXiv:2001.05478 [hep-ph]}}.

\bibitem{Gao:2020vdv}
C.~Gao, J.~Isaacson, and C.~Krause, {\it {i-flow: High-dimensional Integration
  and Sampling with Normalizing Flows}},
\href{http://arxiv.org/abs/2001.05486}{{arXiv:2001.05486 [physics.comp-ph]}}.

\bibitem{Gao:2020zvv}
C.~Gao, S.~Höche, J.~Isaacson, C.~Krause, and H.~Schulz, {\it {Event
  Generation with Normalizing Flows}},
  \href{http://dx.doi.org/10.1103/PhysRevD.101.076002}{Phys. Rev. D {\bfseries
  101} (2020) 7, 076002},
  \href{http://arxiv.org/abs/2001.10028}{{arXiv:2001.10028 [hep-ph]}}.

\bibitem{Chen:2020nfb}
I.-K. Chen, M.~D. Klimek, and M.~Perelstein, {\it {Improved Neural Network
  Monte Carlo Simulation}},
  \href{http://arxiv.org/abs/2009.07819}{{arXiv:2009.07819 [hep-ph]}}.

\bibitem{Verheyen:2020bjw}
R.~Verheyen and B.~Stienen, {\it {Phase Space Sampling and Inference from
  Weighted Events with Autoregressive Flows}},
  \href{http://arxiv.org/abs/2011.13445}{{arXiv:2011.13445 [hep-ph]}}.

\bibitem{Nachman:2020lpy}
B.~Nachman and D.~Shih, {\it {Anomaly Detection with Density Estimation}},
\href{http://arxiv.org/abs/2001.04990}{{arXiv:2001.04990 [hep-ph]}}.

\bibitem{Bellagente:2020piv}
M.~Bellagente, A.~Butter, G.~Kasieczka, T.~Plehn, A.~Rousselot,
  R.~Winterhalder, L.~Ardizzone, and U.~K\"othe, {\it {Invertible Networks or
  Partons to Detector and Back Again}},
  \href{http://arxiv.org/abs/2006.06685}{{arXiv:2006.06685 [hep-ph]}}.

\bibitem{Brehmer:2020vwc}
J.~Brehmer and K.~Cranmer, {\it {Flows for simultaneous manifold learning and
  density estimation}},
  \href{http://arxiv.org/abs/2003.13913}{{arXiv:2003.13913 [stat.ML]}}.

\bibitem{Bothmann:2019yzt}
E.~Bothmann {\em et al.}, {\it {Event Generation with Sherpa 2.2}},
  \href{http://dx.doi.org/10.21468/SciPostPhys.7.3.034}{SciPost Phys.
  {\bfseries 7} (2019) 3, 034},
  \href{http://arxiv.org/abs/1905.09127}{{arXiv:1905.09127 [hep-ph]}}.

\bibitem{coupling1}
L.~Dinh, D.~Krueger, and Y.~Bengio, {\it Nice: Non-linear independent
  components estimation},
  \href{http://arxiv.org/abs/1410.8516}{{arXiv:1410.8516 [cs.LG]}}.

\bibitem{Hoche:2014rga}
S.~H\"oche, \href{http://dx.doi.org/10.1142/9789814678766_0005}{{\it
  {Introduction to parton-shower event generators}}, } in {\em {Theoretical
  Advanced Study Institute in Elementary Particle Physics}: {Journeys Through
  the Precision Frontier: Amplitudes for Colliders}}.
\newblock 2015.
\newblock \href{http://arxiv.org/abs/1411.4085}{{arXiv:1411.4085 [hep-ph]}}.

\bibitem{Plehn:2015dqa}
T.~Plehn, \href{http://dx.doi.org/10.1007/978-3-319-05942-6}{{\em {Lectures on
  LHC Physics}}}, vol.~886.
\newblock Springer, 2015.
\newblock
\newblock
  \href{https://www.thphys.uni-heidelberg.de/~plehn/?visible=review}{\href{http://arxiv.org/abs/0910.4182}{{arXiv:0910.4182
  [hep-ph]}}}.

\bibitem{Catani:1996jh}
S.~Catani and M.~Seymour, {\it {The Dipole formalism for the calculation of QCD
  jet cross-sections at next-to-leading order}},
  \href{http://dx.doi.org/10.1016/0370-2693(96)00425-X}{Phys. Lett. B
  {\bfseries 378} (1996)  287},
  \href{http://arxiv.org/abs/hep-ph/9602277}{{arXiv:hep-ph/9602277}}.

\bibitem{Catani:1990rr}
S.~Catani, B.~R. Webber, and G.~Marchesini, {\it {QCD coherent branching and
  semiinclusive processes at large $x$}},
\href{http://inspirehep.net/search?p=f+j+NUPHA,B349,635}{\href{http://dx.doi.org/10.1016/0550-3213(91)90390-J}{Nucl.
  Phys. {\bfseries B349} (1991)  635}}.

\bibitem{adam}
D.~P. {Kingma} and J.~{Ba}, {\it {Adam: A Method for Stochastic Optimization}},
   \href{http://arxiv.org/abs/1412.6980}{{arXiv:1412.6980 [cs.LG]}}.

\bibitem{Dokshitzer:1997in}
Y.~L. Dokshitzer, G.~Leder, S.~Moretti, and B.~Webber, {\it {Better jet
  clustering algorithms}},
  \href{http://dx.doi.org/10.1088/1126-6708/1997/08/001}{JHEP {\bfseries 08}
  (1997)  001},
  \href{http://arxiv.org/abs/hep-ph/9707323}{{arXiv:hep-ph/9707323}}.

\bibitem{fastjet}
M.~Cacciari, G.~P. Salam, and G.~Soyez, {\it {FastJet User Manual}},
  \href{http://dx.doi.org/10.1140/epjc/s10052-012-1896-2}{Eur. Phys. J. C
  {\bfseries 72} (2012)  1896},
  \href{http://arxiv.org/abs/1111.6097}{{arXiv:1111.6097 [hep-ph]}}.

\bibitem{Kasieczka:2018lwf}
G.~Kasieczka, N.~Kiefer, T.~Plehn, and J.~M. Thompson, {\it {Quark-Gluon
  Tagging: Machine Learning vs Detector}},
  \href{http://dx.doi.org/10.21468/SciPostPhys.6.6.069}{SciPost Phys.
  {\bfseries 6} (2019) 6, 069},
  \href{http://arxiv.org/abs/1812.09223}{{arXiv:1812.09223 [hep-ph]}}.

\bibitem{Frye:2017yrw}
C.~Frye, A.~J. Larkoski, J.~Thaler, and K.~Zhou, {\it {Casimir Meets Poisson:
  Improved Quark/Gluon Discrimination with Counting Observables}},
  \href{http://dx.doi.org/10.1007/JHEP09(2017)083}{JHEP {\bfseries 09} (2017)
  083}, \href{http://arxiv.org/abs/1704.06266}{{arXiv:1704.06266 [hep-ph]}}.

\bibitem{Gallicchio:2010dq}
J.~Gallicchio, J.~Huth, M.~Kagan, M.~D. Schwartz, K.~Black, and B.~Tweedie,
  {\it {Multivariate discrimination and the Higgs + W/Z search}},
  \href{http://dx.doi.org/10.1007/JHEP04(2011)069}{JHEP {\bfseries 04} (2011)
  069}, \href{http://arxiv.org/abs/1010.3698}{{arXiv:1010.3698 [hep-ph]}}.

\bibitem{CMS:2013kfa}
CMS,  \href{http://cds.cern.ch/record/1599732}{{\it {Performance of quark/gluon
  discrimination in 8 TeV pp data}}, }.

\bibitem{Larkoski:2013eya}
A.~J. Larkoski, G.~P. Salam, and J.~Thaler, {\it {Energy Correlation Functions
  for Jet Substructure}},
  \href{http://dx.doi.org/10.1007/JHEP06(2013)108}{JHEP {\bfseries 06} (2013)
  108}, \href{http://arxiv.org/abs/1305.0007}{{arXiv:1305.0007 [hep-ph]}}.

\bibitem{Pumplin:1991kc}
J.~Pumplin, {\it {How to tell quark jets from gluon jets}},
  \href{http://dx.doi.org/10.1103/PhysRevD.44.2025}{Phys. Rev. D {\bfseries 44}
  (1991)  2025}.

\bibitem{deFavereau:2013fsa}
DELPHES 3, J.~de~Favereau, C.~Delaere, P.~Demin, A.~Giammanco, V.~Lemaître,
  A.~Mertens, and M.~Selvaggi, {\it {DELPHES 3, A modular framework for fast
  simulation of a generic collider experiment}},
  \href{http://dx.doi.org/10.1007/JHEP02(2014)057}{JHEP {\bfseries 02} (2014)
  057},
\href{http://arxiv.org/abs/1307.6346}{{arXiv:1307.6346 [hep-ex]}}.

\bibitem{Cacciari:2011ma}
M.~Cacciari, G.~P. Salam, and G.~Soyez, {\it {FastJet User Manual}},
  \href{http://dx.doi.org/10.1140/epjc/s10052-012-1896-2}{Eur. Phys. J.
  {\bfseries C72} (2012)  1896},
\href{http://arxiv.org/abs/1111.6097}{{arXiv:1111.6097 [hep-ph]}}.

\end{thebibliography}\endgroup

\end{document}